\documentclass[aps, prd, amsmath, floats, eqsecnum, superscriptaddress, floatfix, onecolumn, nofootinbib,10pt]{revtex4-2}
\usepackage{graphicx}
\usepackage[format=plain, font=small, labelfont=bf, justification=RaggedRight]{caption}
\usepackage[format=plain, font=small, labelfont=bf, justification=RaggedRight]{subcaption}
\usepackage{color}
\usepackage{latexsym}
\usepackage{slashed}
\usepackage{amssymb}
\usepackage{multirow}

\renewcommand{\thesection}{\arabic{section}}
\renewcommand{\thesubsection}{\thesection.\arabic{subsection}}

\makeatletter
\def\p@subsection{}
\def\p@subsubsection{}
\makeatother

\newcommand{\rd}{\mathrm{d}}	
\usepackage{mathrsfs}			

\newcommand{\bea}{\begin{eqnarray}}
\newcommand{\eea}{\end{eqnarray}}

\newcommand{\beq}{\begin{equation}}
\newcommand{\eeq}{\end{equation}}

\renewcommand{\Re}{\,\mbox{Re}\,}

\newcommand{\eq}[1]{(\ref{#1})}
\newcommand{\n}[1]{\label{#1}}

\newcommand{\al}{\alpha}
\newcommand{\be}{\beta}

\newcommand\De{\Delta}

\renewcommand\th{\theta}

\newcommand\ka{\kappa}

\newcommand\si{\sigma}

\newcommand\ph{\varphi}

\begin{document}

\title{Regular multi-horizon Lee--Wick black holes}

\author{Nicol\`o Burzill\`a}
\email{nburzilla@outlook.it}
\affiliation{
{\small Department of Physics, Southern University of Science and Technology, Shenzhen 518055, China}
}

\author{Breno L. Giacchini}
\email{bgiacchini@gmail.com}
\address{
{\small Departamento de F\'{\i}sica,  ICE, Universidade Federal de Juiz de Fora,
Juiz de Fora,  36036-900,  MG,  Brazil}
}

\author{Tib\'{e}rio de Paula Netto}
\email{tiberiodepaulanetto@gmail.com}
\address{
{\small Departamento de F\'{\i}sica,  ICE, Universidade Federal de Juiz de Fora,
Juiz de Fora,  36036-900,  MG,  Brazil}
}

\author{Leonardo Modesto}
\email{lmodesto@sustech.edu.cn
\vspace{0.05cm}
\\
{\scriptsize This is the Accepted Manuscript version of an article accepted for publication in Journal of Cosmology and Astroparticle Physics. Neither SISSA Medialab Srl nor IOP Publishing Ltd is responsible for any errors or omissions in this version of the manuscript or any version derived from it.  The Version of Record is available online at \url{https://doi.org/10.1088/1475-7516/2023/11/067}.}}
\affiliation{
{\small Department of Physics, Southern University of Science and Technology, Shenzhen 518055, China}
}


\begin{abstract} \noindent
{\bf Abstract}. In this paper we carry out a detailed analysis of the static spherically symmetric solutions of a sixth-derivative Lee--Wick gravity model in the effective delta source approximation. Previous studies of these solutions have only considered the particular case in which the real and the imaginary part of the Lee--Wick mass $\mu=a +i b$ are equal.  However, as we show here, the solutions exhibit an interesting structure when the full parameter space is considered, owing to the oscillations of the metric that depend on the ratio $b/a$. Such oscillations can generate a rich structure of horizons, a sequence of mass gaps and the existence of multiple regimes for black hole sizes (horizon position gaps). In what concerns the thermodynamics of these objects, the oscillation of the Hawking temperature determines the presence of multiple mass scales for the remnants of the evaporation process and may permit the existence of cold black holes with zero Hawking temperature~$T$ and quasi-stable intermediate configurations with $T \approx 0$ and a long evaporation lifetime. For the sake of generality, we consider two families of solutions, one with a trivial shift function and the other with a non-trivial one (dirty black hole). The latter solution has the advantage of reproducing the modified Newtonian-limit metric of Lee--Wick gravity for small and large values of~$r$.
\end{abstract}

\maketitle
\noindent


\vspace{-0.2cm}

\section{Introduction}
\label{Sec1}

From the classical point of view, Lee--Wick gravity is the particular class of higher-derivative gravity models in which some (or all) of the massive degrees of freedom have complex masses. These gravity models were first presented in the works~\cite{ModestoShapiro16,Modesto16} as viable candidates for a consistent theory of quantum gravity. Indeed, while higher derivatives improve the convergence of loop integrals, making the theory super-renormalizable~\cite{AsoreyLopezShapiro}, 
it is possible to preserve the unitarity of the $S$-matrix
if the ghost-like poles of the propagator are complex --- in a similar way as proposed by Lee and Wick~\cite{LW1,LW2} (see also~\cite{CLOP,AnselmiPiva2,Liu:2022gun}).

The replacement of the ghost-like poles that exist in the propagator of fourth-derivative gravity~\cite{Stelle77} by  pairs of complex conjugate poles requires an action with at least six derivatives of the metric. 
In this work we consider the Lee--Wick gravity model described by the action
\beq
\label{action}
S = \frac{1}{16 \pi G} \int \rd^4 x \sqrt{-g} \left[ R + G_{\mu\nu} \left( \al_1 + \al_2 \Box \right)  R^{\mu\nu} \right]
,
\eeq
where $G_{\mu\nu}$ is the Einstein tensor and the parameters $\al_1$ and $\al_2$ are chosen in such a way that the propagator only has complex massive poles. The gauge-independent part of the propagator associated with~\eq{action} 
is given by
\beq
D_{\mu\nu,\al\be}(k)= \frac{1}{k^2 f(-k^2)} \left[  {P}^{(2)}_{\mu\nu,\al\be}(k) - \frac12 {P}^{(0-s)}_{\mu\nu,\al\be}(k) \right] ,
\eeq
where we defined the polynomial function
\beq
\n{fzinho}
f(z) = 1 + \al_1 z + \al_2 z^2 
\eeq
and ${P}^{(2,0-s)}_{\mu\nu,\al\be}$ are the Barnes--Rivers projectors of spin-$2$ and spin-$0$ components~\cite{Barnes,Rivers}. Therefore, the propagator has complex poles provided that the polynomial~\eq{fzinho} 
has a pair of distinct complex conjugate roots, $z = \mu^2$ and $z = \bar{\mu}^2$. 
More general models can include other curvature-squared terms in the action, such as $R\Box^n R$ and $R_{\mu\nu}\Box^n R^{\mu\nu}$ (which increase the number of poles in the propagator) and also terms of higher order in curvatures (which do not change the spectrum, but affect the vertices and the classical equations of motion).

Owing to the difficulties posed by the higher derivatives, the classical solutions obtained in the literature on Lee--Wick gravity involve approximations. 
At linear level, for instance, it was shown that the modified Newtonian potential is a real quantity, even though the spectrum of these models contains particles with complex masses~\cite{Newton-BLG,BreTib1}. Moreover, the complex poles generate oscillatory contributions to the gravitational force, which have been investigated in several low-energy manifestations~\cite{Accioly:2016qeb,Accioly:2016etf,Giacchini:2018twk,BreTib1,Boos:2018bhd,BreLuc}.\footnote{See also~\cite{Krishak:2020opb,Antoniou:2017mhs,Perivolaropoulos:2016ucs} for the analysis of laboratory tests concerning an oscillating gravitational potential.} The Newtonian-limit metric associated with a point-like mass in the model~\eqref{action} (and also in its higher-derivative generalizations) is regular and has regular curvature invariants~\cite{BreTib1}, while invariants containing covariant derivatives of the curvatures might be singular~\cite{Nos6der}.

At nonlinear level, a static and spherically symmetric solution of approximated equations of motion has been obtained for the model~\eq{action} with the restriction $\al_1 = 0$~\cite{Bambi:2016wmo}. 
Different aspects of this solution were studied, such as the curvature regularity and black hole thermodynamics~\cite{Bambi:2016wmo}, the gravitational light deflection~\cite{Zhao:2017jmv,Zhu:2020wtp}, precession of orbits~\cite{Lin:2022wda} and a rotating form following the Newman--Janis algorithm~\cite{Singh:2022tlo}. 
Like in other regular metrics (see, \textit{e.g.},~\cite{FroVilkMG,Hayward:2005gi,Modesto:2010uh,Modesto12,Frolov:Exp,Frolov:Poly,BreTib1,NosG}), there exists a mass gap for the solution to describe a black hole. In other words, the solution does not have an horizon if the mass is smaller than a certain critical value, whereas for larger values of the mass the solution has two horizons and, therefore, it is called ``Lee--Wick black hole''~\cite{Bambi:2016wmo}.

However, the choice $\al_1 = 0$ considerably restricts the space of solutions, since it fixes the ratio between the real and the imaginary part of the Lee--Wick mass $\mu$. As we show in the present work, for nontrivial values of the parameter $\al_1$ the solutions exhibit a rich structure of horizons, oscillations 
and black hole configurations, depending on the scaling of the parameters and the mass of the source. 
Such properties affect also the thermodynamics of these objects, \textit{e.g.}, there are ranges of the parameters $\al_1$ and $\al_2$ such that the solution can describe a cold black hole with Hawking temperature $T \approx 0$, which evaporates slowly and characterises an intermediate configuration with long lifetime.
These interesting features might open the way to new applications and perspectives for testing the model with observational data.

This paper is organized as follows. 
In Sec.~\ref{Sec1.5} we discuss the possible ranges for the parameters $\al_1$ and $\al_2$ in the action~\eqref{action} and how they relate to the Lee--Wick mass $\mu$.
In Sec.~\ref{Sec2} we obtain a Lee--Wick black hole solution for the general model~\eq{action} with nontrivial parameters $\al_1$ and $\al_2$. The solution can be regarded as a three-parameter extension of the two-parameter one obtained in~\cite{Bambi:2016wmo}. We show that the solution is regular and discuss the structure of horizons depending on the values of the parameters in the model. 
In Sec.~\ref{Sec3} we present another family of Lee--Wick black holes, which has the advantage of reproducing the Newtonian-limit metric in the regimes of large and small $r$. These solutions are also regular and have the same horizon structure of the ones discussed in the preceding section.
The evaporation of Lee--Wick black holes is studied in Sec.~\ref{Sec5}, where we show that the possibilities for the mass of the remnant depend on the structure of horizons; moreover, there can be cold black hole  configurations with a long lifetime.
In Sec.~\ref{Sec6} we summarize the results and draw our conclusions. 
Technical details are presented in three Appendices.
Finally, throughout this paper we use the same sign conventions of~\cite{BreTibLiv} and we adopt the unit system such that $c = 1$ and $\hslash  = 1$.


\section{Parameter space of Lee--Wick sixth-derivative gravity}
\label{Sec1.5}

Before considering the black hole solutions, it is instructive to determine the possible range of the parameters $\al_{1,2}$ in the action~\eq{action} and their relation to other quantities we use along this work.
Writing the polynomial~\eq{fzinho} in the factored form
\beq
\label{fsixd}
f(z) = \frac{(\mu^2 - z) (\bar{\mu}^2 - z) }{\vert\mu\vert^4} , 
\eeq
and defining the Lee--Wick mass $\mu$ in terms of its real and imaginary parts,\footnote{Notice that all the quantities in this section only depend on $a^2$ and $b^2$, which makes the choice $a,b>0$ look somewhat restrictive at this stage. This requirement is motivated by the next section, where we select the square root of $\mu^2$ with positive real part $a>0$ in order to meet the boundary conditions for having an effective source that vanishes for $r\to\infty$, and an asymptotically flat metric, as discussed also in~\cite{Accioly:2016qeb}. On the other hand, there is no loss of generality in $b>0$, inasmuch as $\mu$ and $\bar{\mu}$ have imaginary parts with opposite signs.} 
\beq
\label{Defab}
\mu \equiv a + i b , \qquad a,b>0, 
\eeq
it is straightforward to obtain the relation between $\mu$ and the parameters in the action~\eq{action},
\beq
\label{Rel1}
\alpha_1 = - \frac{2\Re(\mu^2)}{|\mu|^4} = - \frac{2 (a^2-b^2)}{(a^2+b^2)^2} \, ,
\qquad
\alpha_2 = \frac{1}{|\mu|^4} = \frac{1}{(a^2+b^2)^2 }  \, .
\eeq

The relations~\eq{Rel1} can be inverted, namely,
\beq
\label{Rel2}
a^2 = \frac{2 \sqrt{\alpha_2} - \alpha_1}{4 \alpha_2} ,
\qquad 
b^2 = \frac{2 \sqrt{\alpha_2} + \alpha_1}{4 \alpha_2} ,
\eeq
which results in the following restriction to the parameters of the action:
\beq
\alpha_2  > 0 
\qquad \text{and} \qquad
-2 \sqrt{\alpha_2} < \alpha_1  < 2 \sqrt{\alpha_2}.
\eeq
Notice that the last inequalities must be strict in order to guarantee that $\mu$ is neither tachyonic (\textit{i.e.}, $a^2 \neq 0$) nor has a vanishing imaginary part (\textit{i.e.}, $b^2 \neq 0$).
The possibility of having $\al_1 > 0$ is a distinctive feature of Lee--Wick gravity models~\cite{Accioly:2016qeb} and it yields $b > a$ [see Eq.~\eq{Rel2}].

Finally, from the previous equations it is easy to see that the choice $\al_1=0$ 
corresponds to fixing $a=b$. As mentioned in the Introduction, this was the particular choice of parameters adopted in~\cite{Bambi:2016wmo}. 


\section{Lee--Wick black holes}
\label{Sec2}

The first type of Lee--Wick black holes we consider in this paper is the direct extension of the solutions obtained in~\cite{Bambi:2016wmo} to the case of a non-trivial parameter $\al_1$ in the action~\eq{action}. Such solutions are described by a Schwarzschild-like metric in the form
\beq
\label{metricB=0}
\rd s^2 = - A(r)  \rd t^2 + \frac{\rd r^2}{A(r)} + r^2 \rd \Omega^2, 
\eeq
which solves the effective field equations
\beq
\label{Gff}
G^{\mu} {}_{\nu} = 8 \pi G \, \tilde{T}^{\mu} {}_{\nu},
\eeq
where 
\beq
\n{effT}
\tilde{T}^{\mu} {}_\nu = \text{diag}(-\rho , p_r , p_\th , p_\th ) 
\eeq
is an effective energy-momentum tensor. In Eq.~\eq{effT}, the effective source $\rho(r)$ is the smeared delta source~\cite{BreTib2,Bambi:2016wmo,BreTibLiv}, namely,
\beq
\begin{split}
\n{effsource}
\rho(r) & =  \frac{M }{2 \pi ^2 r} \int_0^\infty \rd k \, \frac{ k \sin (kr)}{ f(-k^2) } =\frac{M \vert\mu\vert^4}{2 \pi ^2 r} \int_0^\infty \rd k \, \frac{ k \sin (kr)}{ (\mu^2 + k^2) (\bar{\mu}^2 + k^2) }
\\
& =  \frac{M (a^2+b^2)^2}{8 \pi ab } \frac{e^{-ar}\sin(b r) }{r}
,
\end{split}
\eeq
where we used~\eq{fsixd} and~\eq{Defab}. 
On the other hand, the effective pressure components $p_r$ and $p_\th$ can be determined by the field equations together with the conservation equation $\nabla_\mu \tilde{T}^\mu {}_\nu = 0$, similarly to the procedure of~\cite{Bambi:2016wmo,Nicolini:2005vd,Modesto12,Modesto:2010uh}. 
The role of the effective energy-momentum tensor~\eq{effT} is to mimic the effect of the higher derivatives and compensate the truncation of the complete equations of motion originated from~\eq{action}, 
which read\footnote{The terms of higher order in curvature are collectively denoted by ${O}({R}^2_{\ldots})$; the explicit expression can be found, \textit{e.g.}, in~\cite{Accioly:2016qeb,Decanini:2007gj}.}
\beq
\n{EOMc}
\left( 1 + \al_1 \Box + \al_2 \Box^2 \right) G^{\mu} {}_{\nu} + {O}({R}^2_{\ldots}) = 8 \pi G \, T^{\mu} {}_{\nu}.
\eeq 
For further details, see~\cite{Bambi:2016wmo,NosG}.

Finally, in terms of the mass function $m(r)$, defined such that
\beq
\label{Am}
A(r) = 1 - \frac{2G {m}(r)}{r},
\eeq
the field equations~\eq{Gff} are equivalent to the system
\begin{subequations} \label{System}
\begin{align} 
 \n{EqGtt}
\frac{\rd {m}}{\rd r} & =  4 \pi r^2 \rho \,=\, - 4 \pi r^2 p_r
\,, 
\\
 \n{Conserva}
\frac{\rd p_r}{\rd r}  & =   \frac{2}{r} \left( p_\th - p_r \right) - \frac{\left( p_r + \rho \right)}{2A} \frac{\rd A}{\rd r} .
\end{align}
\end{subequations}
The system can be solved for the effective pressures and the mass function, resulting in 
\beq
p_r(r) = - \rho(r), \qquad\quad p_\th(r) = - \rho(r) - \frac{r }{2} \rho^\prime(r),
\eeq
and
\beq
\n{meff}
\begin{split}
{m}(r) & =  4\pi \int_0^r \rd x \, x^2 \rho (x) 
\\
& =  M-\frac{M}{2 a b } e^{-a r} \Big\{ b \left[  2a +(a^2+b^2)r \right] \cos(b r) 
+ \left[ a^2-b^2 +a(a^2+b^2)r \right]  \sin(b r) \Big\}
.
\end{split}
\eeq


\subsection{Oscillations of the mass function}
\label{Sec2.massfunc}

The above discussion reveals that the central quantities defining the solution are the effective delta source~\eq{effsource} and the associated mass function~\eq{meff}. General results on the behavior of these functions for generic higher-derivative gravity models  in the regimes of large and small $r$ have been studied in~\cite{BreTib2,Nos6der} (see also~\cite{BreTibLiv} for a didactic introduction). In the case considered here, it is straightforward to obtain 
\beq
\label{mrBIG}
\lim_{r\to\infty} m(r) =  M
,
\eeq
showing that the solution approaches Schwarzschild as $r\to\infty$, while around $r=0$ we have
\beq
\label{mrsmall}
m(r) = \frac{M (a^2+b^2)^2}{6 a} r^3  + O(r^4).
\eeq
In Sec.~\ref{Sec2.reg} below we show that this small-$r$ behavior guarantees the regularity of the solution.

As mentioned in the Introduction, the oscillating contributions to the gravitational potential, effective source and mass function are among the most characteristic features of gravity models with complex poles in the propagator~\cite{Newton-BLG,Accioly:2016qeb,BreTib2,Bambi:2016wmo}. In this particular case, such oscillatory terms are manifest in Eqs.~\eq{effsource} and~\eq{meff}. The amplitude of these oscillations is closely related to the ratio 
\beq
\label{just_q}
q \equiv \frac{b}{a},
\eeq
between the imaginary and real parts of the Lee--Wick mass $\mu$; indeed, the amplitude increases with $q$.
This can be readily seen by
rewriting the effective source~\eq{effsource} in terms of the dimensionless variable
\beq
\label{y_e_q}
y \equiv b r,  
\eeq
namely, 
\beq
\label{rhotil}
\rho 
=
\frac{M a^3 (1+q^2)^2}{8 \pi } \frac{ e^{-\frac{y}{q}} \sin y}{y}.
\eeq
Similar dependence on $q$ is found for the mass function in~\eq{meff}. In fact, defining the dimensionless mass function $\tilde{m}(y)$ such that
\beq
m(r) = M \tilde{m}(br),
\eeq
it is straightforward to get
\beq
\label{mtil}
\tilde{m}(y)  =  1 - \frac{1}{2 q^2} e^{-\frac{y}{q}} \left\lbrace  q [y+q (2+q y)] \cos y -  \left[q (q^2-1) - (q^2+1) y\right] \sin y  \right\rbrace .
\eeq

\begin{figure}[h]
\includegraphics[width=8cm]{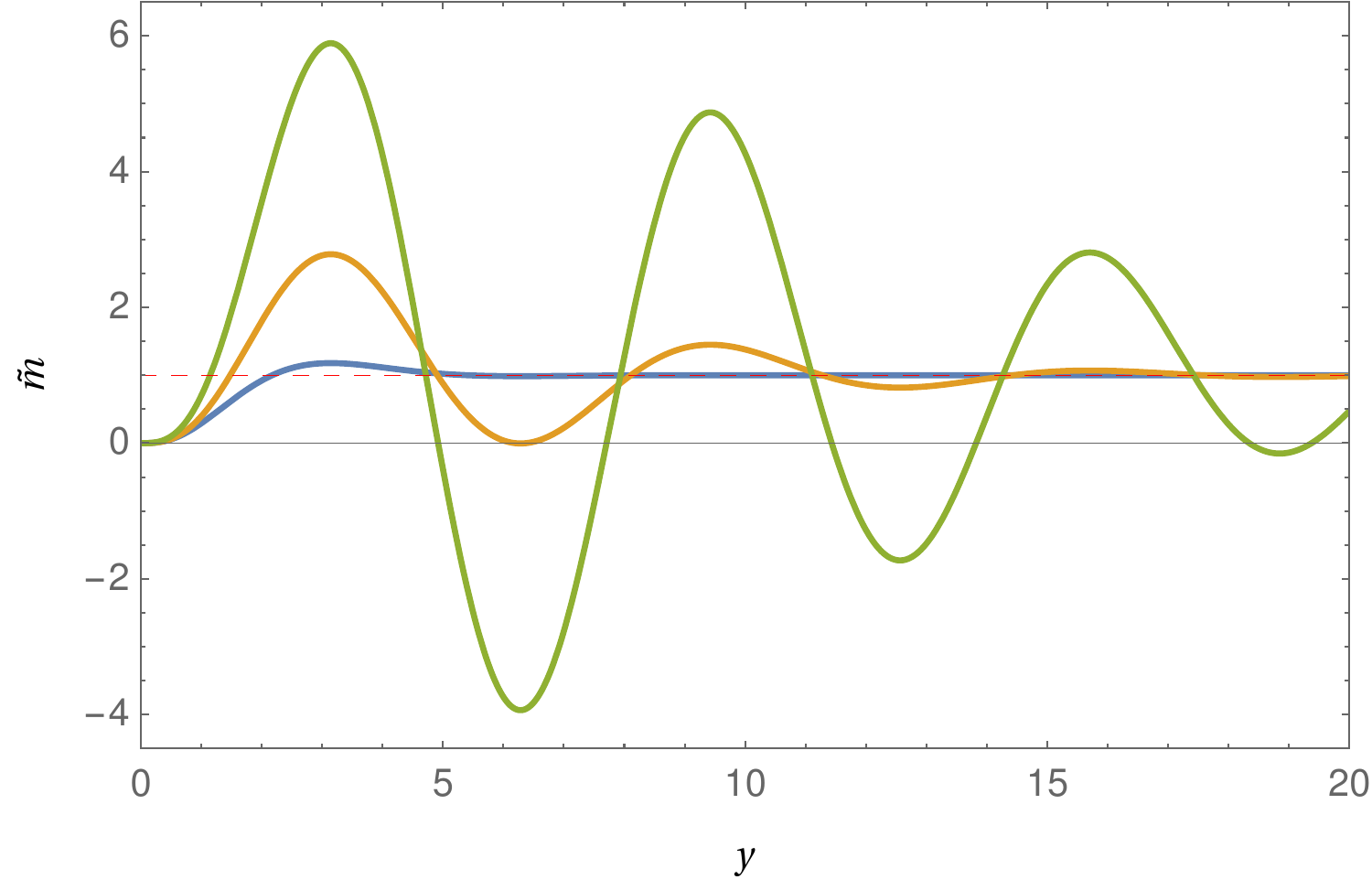} 
\caption{\small 
Graph of the dimensionless effective mass function $\tilde{m}(y)$ for $q=1$ (blue), $q=2.67$ (orange) and $q=5$ (green). The local extrema of $\tilde{m}(y)$ are located at $y = k \pi$ ($k=1,2,\ldots$), and the amplitude of oscillations increases with $q$. In particular, $\tilde{m}(y)$ assumes negative values for $q>2.67$.}
\label{Fig1}
\end{figure}

Since the effective source for Lee--Wick gravity is not strictly positive, the mass function is not monotonic and it can even assume negative values if $q$ is sufficiently large. For instance, inspecting the graph of $\tilde{m}(y)$ for a range of values of $q$, it is straightforward to verify that if $q > 2.67$ there are regions where $\tilde{m}(y) < 0$, see Fig.~\ref{Fig1}. The amplitude of the oscillations can be arbitrarily large as $q$ increases. Such strong oscillation is not present in the case $q=1$ analyzed so far in the literature~\cite{Bambi:2016wmo,Zhao:2017jmv,Zhu:2020wtp,Lin:2022wda} and it opens the possibility for the metric to have more than two horizons, as we discuss in the next section.

Regarding the variation of the amplitude of the oscillations for $q$ fixed, it appears from Fig.~\ref{Fig1} that the oscillation amplitude always decreases with $y$, but this is true only for small values of $q$. As we show in Appendix~\ref{Appendix}, for larger values of $q$ the amplitude of oscillation increases up to a maximum, before decreasing to zero. Moreover, the extrema of $\tilde{m}(y)$ occur at $y = k \pi$ ($k=1,2,\ldots$), as it can be seen in Fig.~\ref{Fig1}.

It is worthwhile to notice that, since for large $y$ the oscillation amplitude decreases and $\tilde{m}(y)\to 1$, there is only a limited number of regions where $\tilde{m}(y)<0$. As we show in what follows, these regions with negative effective mass have an important role in the structure of horizons.

\subsection{Structure of horizons and mass gaps}
\label{Sec2.hor}

\subsubsection{Maximal number of horizons}
\label{maxo}

\begin{figure}[h]
\includegraphics[width=8cm]{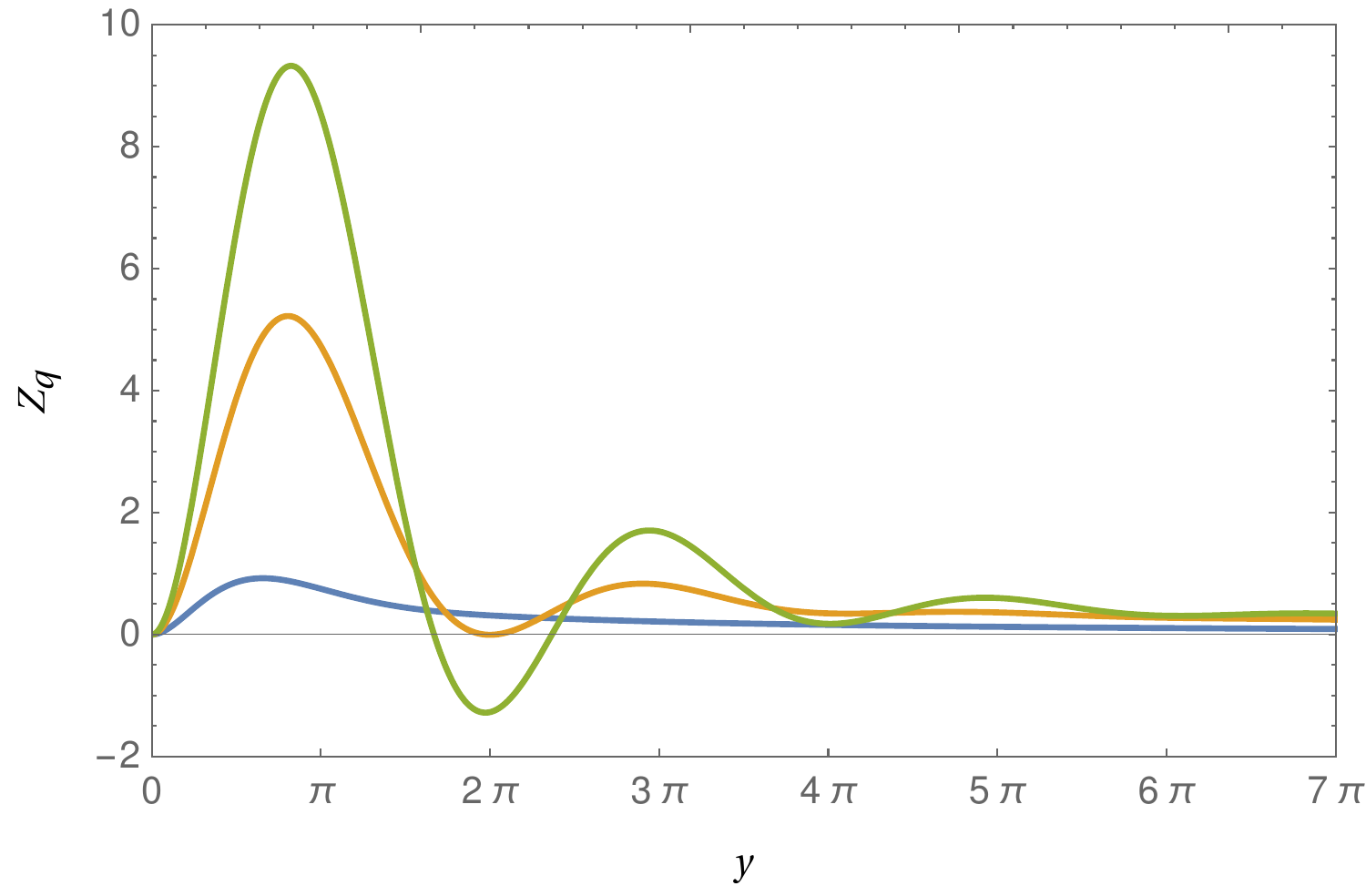} 
\caption{\small Graph of $Z_q(y)$, given by \eqref{Z}, for the cases $q=1$ (blue), $q=2.67$ (orange) and $q=3.5$ (green). Like the dimensionless mass function \eqref{mtil}, the function $Z_q(y)$ can assume negative values for $q>2.67$. The local maxima of $Z_q$ constitute a monotonically decreasing positive sequence.
On the other hand, the sequence of minima is monotonically increasing, but it changes sign from negative to positive if $q>2.67$.}
\label{PlotZ}
\end{figure}

The horizons of the metric~\eq{metricB=0} are defined as the locus of the points where the function $A(r)$ changes sign; therefore, they are related to the zeros of the equation $A(r)=0$. In terms of the definitions~\eq{just_q} and~\eq{y_e_q}, Eq.~\eq{Am} reads
\beq
\label{Az}
A(r) = 1 - G M a \, Z_q(y(r)),
\eeq
where
\beq
\label{Z}
Z_q(y) = 2 q \frac{\tilde{m}(y)}{y} .
\eeq
From the behavior of the mass function [see the discussion related to Eqs.~\eq{mrBIG} and~\eq{mrsmall}], it follows that $Z_q(y)$ is bounded for any\footnote{Remember we always assume $q>0$, see Eq.~\eq{Defab}.} fixed $q$, in particular,
\beq
\label{oslimites}
\lim_{y\to 0} Z_q(y)  = \lim_{y\to \infty} Z_q(y) = 0  ,
\eeq
and it is positive for $y$ sufficiently small or sufficiently large (see Fig.~\ref{PlotZ}). 

Therefore, for every $q$, $Z_q(y)$ must have at least one local maximum but, depending on how~\eq{Z} oscillates, the number $n_\text{E}$ of local extrema can exceed one. Since the number $N_\text{H}$ of horizons of the solution corresponds to the number of times the function $Z_q(y)$ crosses the value $1/(G M a)$ [see Eq.~\eq{Az}], and taking into account that the first extremum of $Z_q(y)$ is always a local maximum, we have the upper bound\footnote{In principle, the function $Z_q(y)$ could even have an infinite number of oscillations;  however, as we show below, this does not happen and $n_\text{E}$ is always limited. In addition, from~\eq{oslimites} and the fact that $Z_q(y)$ is positive in the limits of small and large $y$, it follows that $n_\text{E}$ is odd and, by extension, $N_\text{H}^{\text{max}}$ is even. Note that $ Z_q(y)$ can also have saddle points, which are critical points but are \emph{not} counted as extrema; in fact, the presence of saddle points do not affect the number of horizons.}
\beq
\label{Rel}
N_\text{H} \leqslant n_\text{E} +1 \equiv N_\text{H}^{\text{max}}.
\eeq
Of course, the exact number of horizons will also depend on the value of the mass $M$, but it will not surpass $N_\text{H}^{\text{max}}$.

With this correlation between the maximal number of horizons and the number of local extrema of the function $Z_q(y)$  we can investigate how the scaling of the real and imaginary parts of the Lee--Wick mass affects the distribution of horizons. To this end, let us consider 
\beq
\frac{\partial}{\partial y} Z_q(y)= \frac{1}{q^2 y^2 } \mathcal{G}(y,q),
\eeq
where we defined the function
\beq
\label{G}
\mathcal{G}(y,q)
= -2q^3 +e^{-\frac{y}{q}} \Big[(2q^3+q^2(1+q^2) y)\cos(y)
+(q^2(1-q^2)+q(1+q^2)y+(1+q^2)^2 y^2) \sin(y) \Big] .
\eeq
Hence, the extrema 
of $Z_q(y)$ are the solutions of the equation
\beq
\label{Gzero}
\mathcal{G}(y,q) = 0.
\eeq

\begin{figure}[h]
\includegraphics[scale=0.9]{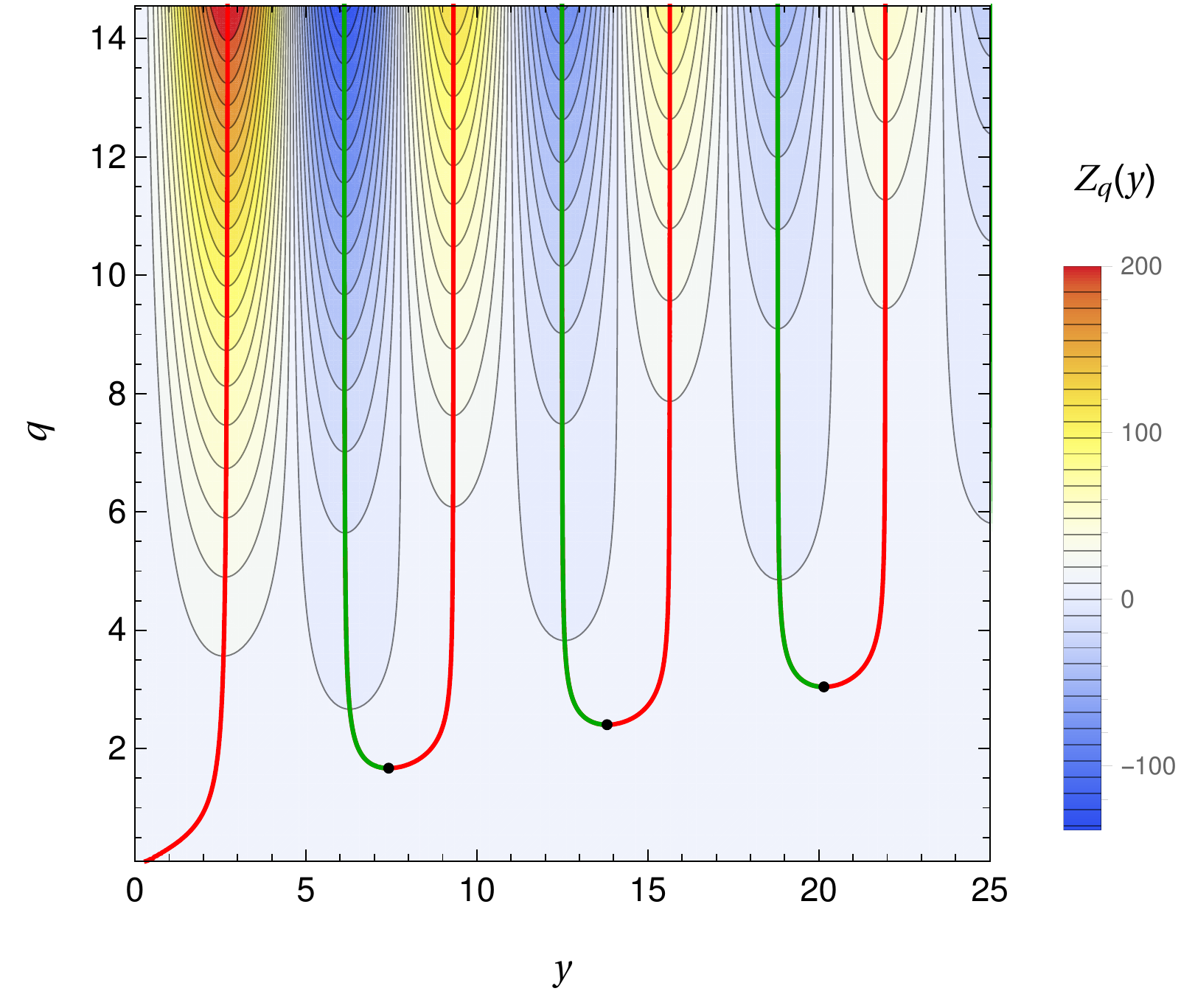}
\caption{\small  Plot of the function $Z_q(y)$ in the $yq$-plane. 
The red curves follow the local maxima of $Z_q(y)$, while the green curves follow the local minima. The black dots in the intersection of these two colors are saddle points. 
Remember that the amplitude of the oscillations of $\tilde{m}(y)$ increases with $q$; this manifests here as the increase of the number and of the amplitude of the oscillations of $Z_q(y)$ for larger values of $q$. Also, the positions of the extrema tend to approach those of $\tilde{m}(y)$, namely, near multiples of $\pi$, for large $q$ (see Appendix~\ref{Appendix} for further details). Finally, by looking at the values of the function $Z_q$, we see that for any $q$ its maxima (minima) constitutes a monotonically decreasing (increasing) sequence. }
\label{Fig2}
\end{figure}

In Fig.~\ref{Fig2} we plot the function $Z_q(y)$ in the $yq$-plane. The solutions of~\eq{Gzero} are represented by the red and green curves. The warm (cool) colors represent the regions where $Z_q(y)$ is positive (negative). Thus, the red (green) curves correspond to the local maxima (mimina) of $Z_q(y)$, while the black dots denote the inflection points along these curves. 

Notice that for each $q$, Eq.~\eq{Gzero} only has a finite number of solutions, which means that the number $n_\text{E}$ of local extrema of $Z_q(y)$ is also finite and odd. Hence, from the relation~\eq{Rel} it follows that for each $q$ there exists a maximal number of horizons the metric can have; moreover, this number grows (in discrete increments) as  $q$ increases.

These discrete increments are related to the saddle points marked by black dots in Fig.~\ref{Fig2}, which coincide with the minima of the red and green curves. Indeed, let us denote the saddle points by $P_\ell = (y^*_\ell, q^*_\ell)$, with $\ell = 1,2,\ldots$, ordered by increasing values of $y^*$. Since there exists a neighborhood $\mathscr{I}_\ell$ around each $P_\ell$ where Eq.~\eq{Gzero} can be solved for $q$, by applying the implicit function theorem it follows that $(y_\ell^*,q_\ell^*)$ are the stationary points of the implicit function $q(y_\ell^*)$. 
The first ten saddle points are listed in Table~\ref{Tab1}. Notice that the values of $q_\ell^*$ also increase with $\ell$ (see Appendix~\ref{Appendix} for a detailed discussion);
thus, if we define $q^*_0 \equiv 0$,  for any $q$ we have (see Fig.~\ref{Fig2})
\beq
q^*_\ell < q < q^*_{\ell+1} \quad \Longrightarrow \quad n_\text{E} = 2\ell + 1,
\eeq
whence, from~\eq{Rel},
\beq
\label{NmaxEll}
 N_\text{H}^{\text{max}} = 2(\ell + 1) .
\eeq
Since the first saddle point has $q_1^*=1.67$, only if $q$ is larger than this value the metric can have more than two horizons.

After a sequence of approximations involving the positions of the saddle points, it is possible to obtain 
an estimate for the maximal number of horizons given a value for the parameter $q\geqslant 1$, namely,
\beq
\label{HmaxAPP}
N_\text{H}^{\text{max}} (q) \, \approx \, 2 \bigg\lceil  - \frac{q}{\pi }  \, W_{-1} \left(-\frac{\sqrt{q}}{\sqrt{2}(q^2-1)}\right) - \frac14 \bigg\rceil   ,
\eeq
where $\lceil x\rceil $ is the ceiling function and $W_{-1}(x)$ denotes the branch of the Lambert $W$ function satisfying $W(x) \leqslant -1$~\cite{LambertW}. The derivation of this result is presented in Appendix~\ref{Appendix} [see Eq.~\eq{NEAPP}]. In Fig.~\ref{Fig5} we show the graph of~\eq{HmaxAPP} in the range $q\in [1,10]$. The approximation character of~\eq{HmaxAPP} is perceived in the error at the position of the transition points between the different maximal numbers of horizons; this error, however, tends to decrease with $q$. For example, according to~\eq{HmaxAPP} the transition from 2 to 4 horizons happens at $q = 1.72$ (while the correct value is $q_1^*=1.67$), and the one from 20 to 22 horizons occurs for $q=6.719$ (the correct value is $q_{10}^*=6.712$).

\begin{minipage}[b]{0.45\textwidth}
    \centering
    \begin{tabular}{|r|r|r|}
    \hline
      $\ell$ & $ q^{*}_\ell $ &  $y^{*}_\ell$ \\
    \hline
    \hline
        1 & 1.670   & 7.420  \\ \hline
        2 & 2.405   & 13.806  \\ \hline
        3 & 3.044 & 20.148   \\ \hline
        4 & 3.634 & 26.470  \\ \hline
        5 & 4.191 & 32.782   \\ \hline
        6 & 4.726 & 39.086   \\ \hline
        7 & 5.242 & 45.386   \\ \hline
        8 & 5.743 & 51.683   \\ \hline
        9 & 6.233 & 57.977   \\ \hline
       10 & 6.712 & 64.270   \\ \hline
    \end{tabular}
\vspace{0.3cm}
\captionof{table}{\small  The first ten saddle points of $Z_q(y)$ along the curve $\mathcal{G}(y,q) = 0$.}
    \label{Tab1}
\vspace{1.1cm}
\end{minipage}
\hfill
\begin{minipage}[b]{0.45\textwidth}
    \centering
    \vspace{0.5cm}
\includegraphics[width=0.75\textwidth]{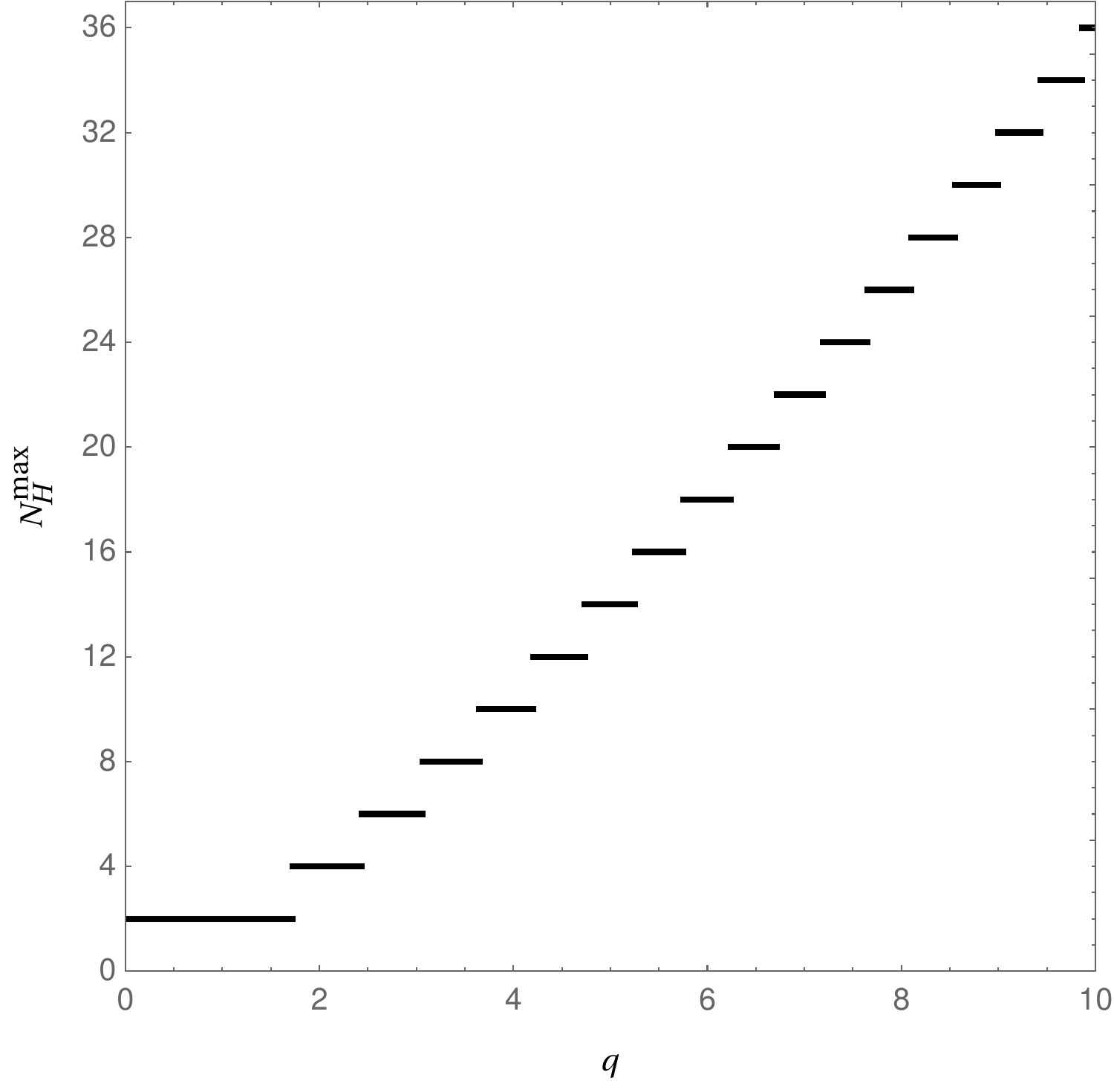}
\captionof{figure}{\small Estimate for the maximal number of horizons $N_\text{H}^{\text{max}}$ using~\eq{HmaxAPP} for $q \in [1,10]$. 
The approximation~\eq{HmaxAPP} cannot be applied for some values in the range $0<q<1$, however, as explained in the main text, in this interval we have exactly $N_\text{H}^{\text{max}}=2$. \label{Fig5}}
    \end{minipage}


\subsubsection{Actual number of horizons}
\label{Sec.ANH}

The actual number of horizons depends on the mass $M$. As explained before, the roots of the equation $A(r) = 0$ are found when the graph of $Z_q(y)$ crosses the horizontal line 
\beq 
\label{LineL}
L(y)= \frac{1}{G M a}
,
\eeq
see Eq.~\eq{Az}.
In what follows we describe the possible situations encountered as one increases the value of $M$.
\begin{enumerate}
\item{
There exists a critical mass 
\beq
M_0 = \frac{1}{G a \max_y\lbrace Z_q(y)\rbrace } 
\eeq
such that if $M < M_0$
the line $L(y)$~\eq{LineL} 
does not cross the graph of $Z_q(y)$ and
the solution has no horizon. Hence, $M_0$ represents the mass gap for the solution to describe a black hole. Such mass gap occurs because $Z_q(y)$ is bounded, and it is an ubiquitous feature of
solutions associated with higher-derivative gravity models~\cite{FroVilkMG,Modesto:2010uh,Modesto12,Frolov:Exp,Frolov:Poly,BreTib1,NosG}. In this particular case, 
owing to the oscillations caused by the Lee--Wick mass function, it is difficult to find an explicit expression for the dependence of $M_0$ on $q$. Nevertheless, recalling that the amplitude of oscillations increases with $q$, one can expect $M_0$ to decrease with $q$.
} 

\item{
As one gradually increases $M > M_0$, the line $L(y)=1/(G M a)$ moves down and crosses the graph of $Z_q(y)$ a number of times. Since the successive maxima of $Z_q(y)$ form a strictly decreasing positive sequence
(see Figs.~\ref{PlotZ} and~\ref{Fig2}), 
there exist some ranges of values of $M$ for which the metric
can have 2, 4, 6, \textit{etc}., horizons. There is also the possibility that the line $L(y)=1/(G M a)$ is only tangent to one of the extrema of $Z_q(y)$. The point where this happens is often referred to as an \emph{extremal horizon}.
}

\item{
The minima of $Z_q(y)$ form a strictly increasing sequence, however, it may happen that some of them are positive (see Figs.~\ref{PlotZ} and~\ref{Fig2}). Therefore, for even larger values of $M$, the quantity $1/(G M a)$ might become smaller than some of the positive minima of $Z_q(y)$. This means that the number of horizons starts to decrease as one increases $M$ beyond a certain value.}

\item{
Finally, there exists another critical value for the mass, which we denote by $M_\text{c}$, such that $M > M_\text{c}$ implies that $1/(G M a)$ is smaller than all the positive minima of $Z_q(y)$ --- an approximate expression for $M_\text{c}$ is provided by Eq.~\eq{EstMc} in Appendix~\ref{Appendix}. As a conclusion, for $M > M_\text{c}$ the metric has a fixed number of horizons. 
Besides the two ``trivial'' horizons (owed to the fact that $Z_q(y)$ is positive in the limits of small and large $y$), 
the other horizons that remain are related to the negative minima of $Z_q(y)$, which, in turn, are originated from the regions where the effective mass is negative [see the discussion following Eq.~\eq{mtil}]. In particular, if $\tilde{m}(y)>0$ for all $y$, then all the minima of $Z_q(y)$ are positive and the metric ends up with two horizons in the limit of large $M$.
}
\end{enumerate}

\begin{table}[b]
    \centering
    \begin{tabular}{|c|c|c|}
            \hline
                & Mass gaps & Number of  \\
                & (in units of $M_{\rm P}^2/a$) & horizons \\
            \hline
            \hline
            \multirow{2}{*}{$q=1$} & \multicolumn{1}{c|}{$M<1.082$} & \multicolumn{1}{c|}{$0$}\\\cline{2-3}
                                 & \multicolumn{1}{c|}{$M>1.082 $} & \multicolumn{1}{c|}{$2$}  \\\hline
            \hline
            \multirow{4}{*}{$q=2$} & \multicolumn{1}{c|}{$M<0.345 $} & \multicolumn{1}{c|}{$0$} \\\cline{2-3}
                                 & \multicolumn{1}{c|}{$0.345 <M<2.024 $} & \multicolumn{1}{c|}{$2$} \\\cline{2-3}
                                 & \multicolumn{1}{c|}{$2.024 <M<2.593 $} & \multicolumn{1}{c|}{$4$} \\\cline{2-3}
                                 & \multicolumn{1}{c|}{$M>2.593 $} & \multicolumn{1}{c|}{$2$} \\\hline                                 
            \hline
            \multirow{6}{*}{$q=3$} & \multicolumn{1}{c|}{$M<0.149 $} & \multicolumn{1}{c|}{$0$} \\\cline{2-3}
                                 & \multicolumn{1}{c|}{$0.149 <M<0.901 $} & \multicolumn{1}{c|}{$2$} \\\cline{2-3}
                                 & \multicolumn{1}{c|}{$0.901 <M<2.254 $} & \multicolumn{1}{c|}{$4$} \\\cline{2-3}
                                 & \multicolumn{1}{c|}{$2.254 <M<3.158 $} & \multicolumn{1}{c|}{$6$} \\\cline{2-3}                  
                                 & \multicolumn{1}{c|}{$M>3.158 $} & \multicolumn{1}{c|}{$4$}  \\\hline                                
        \end{tabular}    
        \caption{\small Ranges of values of $M$ and number of horizons, for three different values of $q$. The parameter $a$ is left arbitrary and $M_{\rm P}^2 = 1/G$ denotes the square of the Planck mass.}
\label{Tab2}
\end{table}

\begin{figure}[h]
\begin{subfigure}{.5\textwidth}
\centering
\includegraphics[width=7cm]{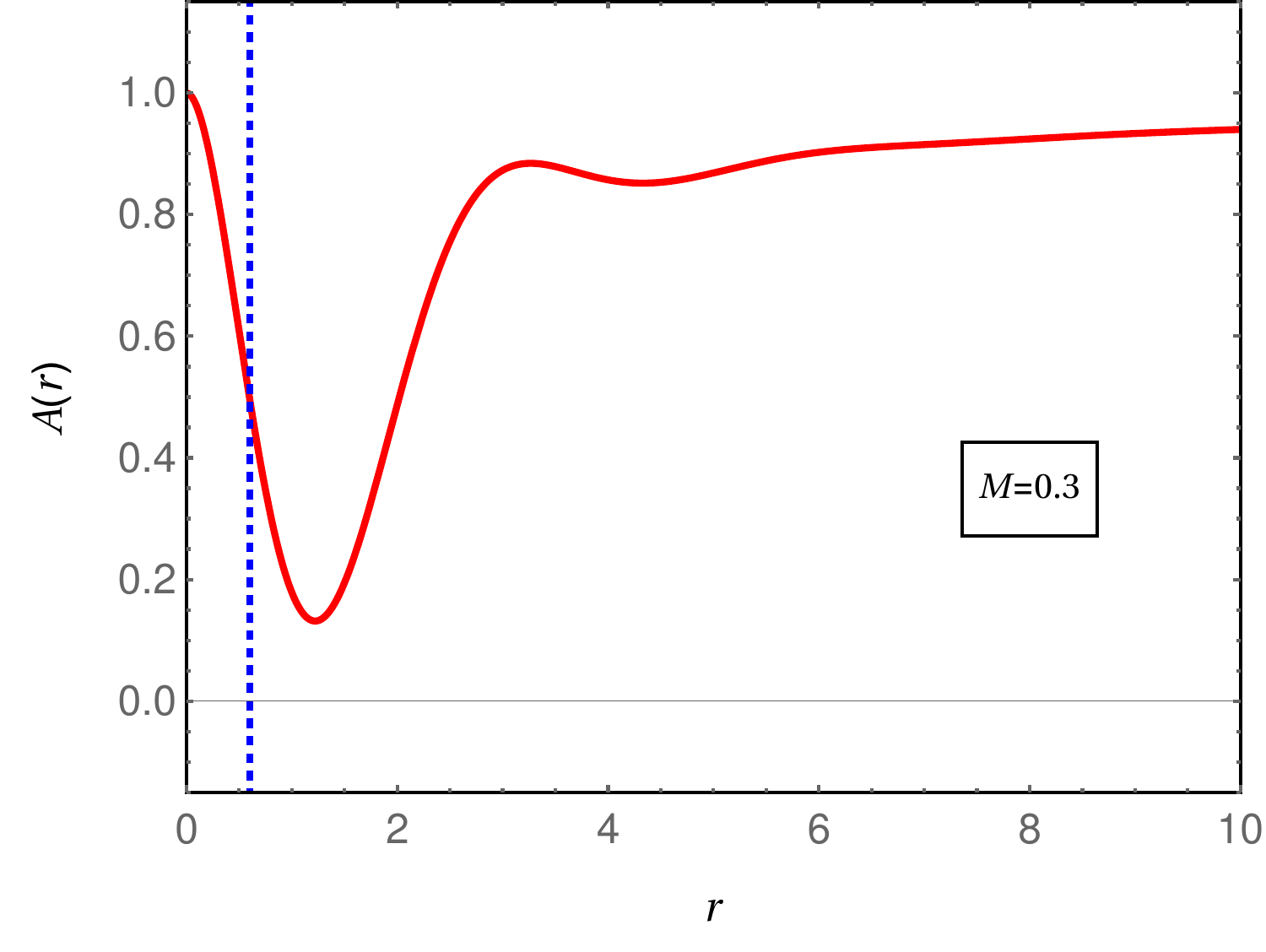}
\end{subfigure}%
\begin{subfigure}{.5\textwidth}
\centering
\includegraphics[width=7cm]{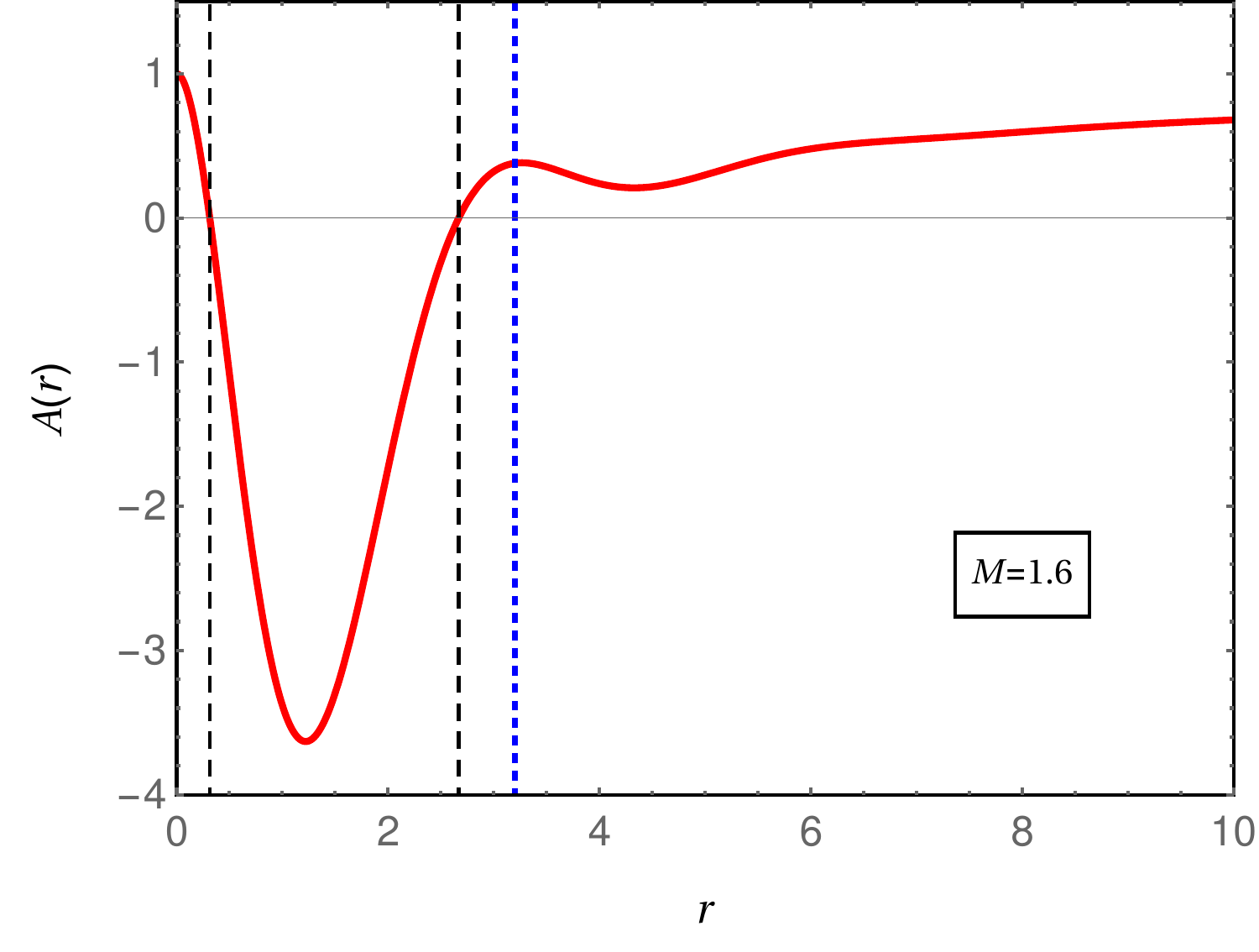}
\end{subfigure}%
\\
\begin{subfigure}{.5\textwidth}
\centering
\includegraphics[width=7cm]{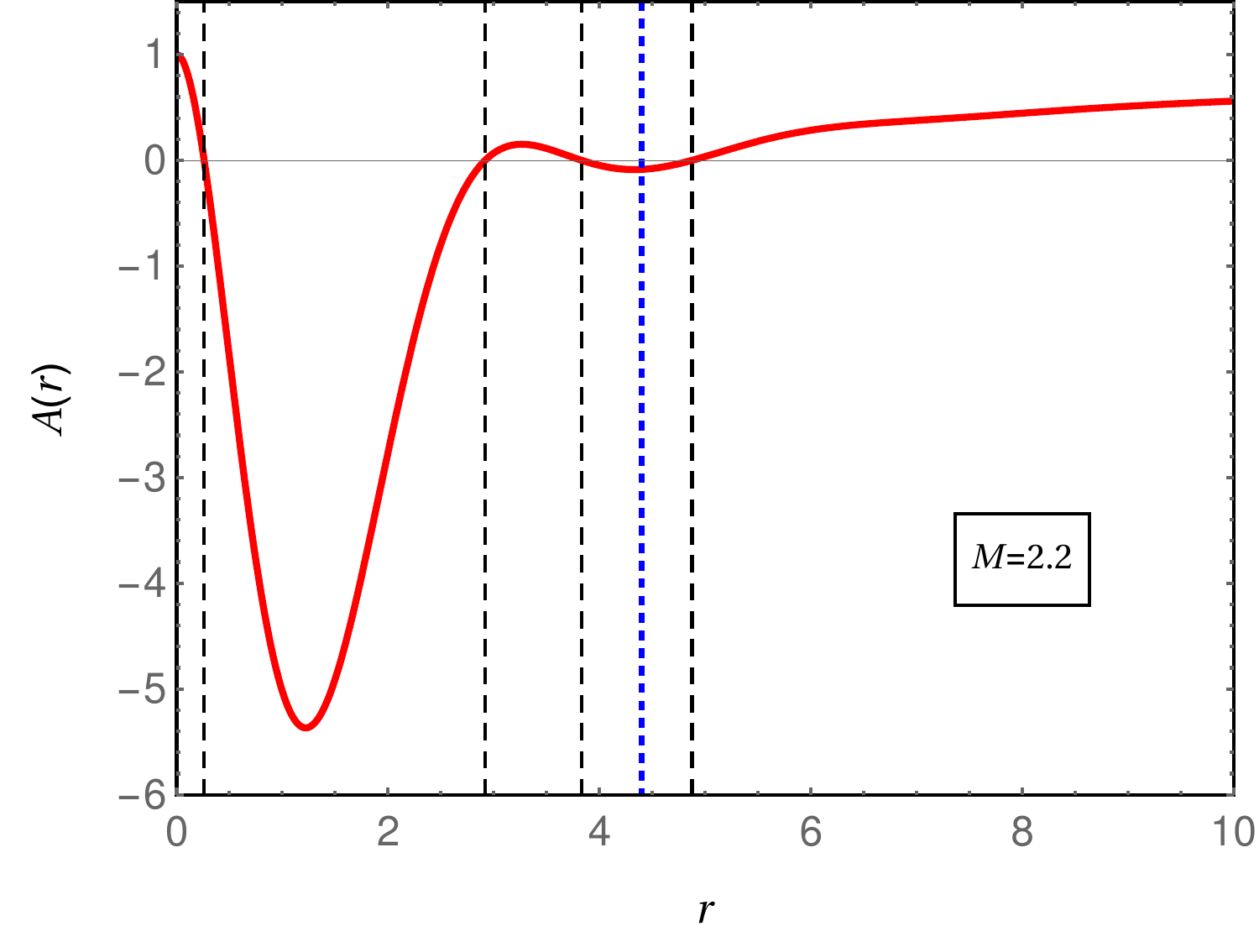}
\end{subfigure}%
\begin{subfigure}{.5\textwidth}
\centering
\includegraphics[width=7cm]{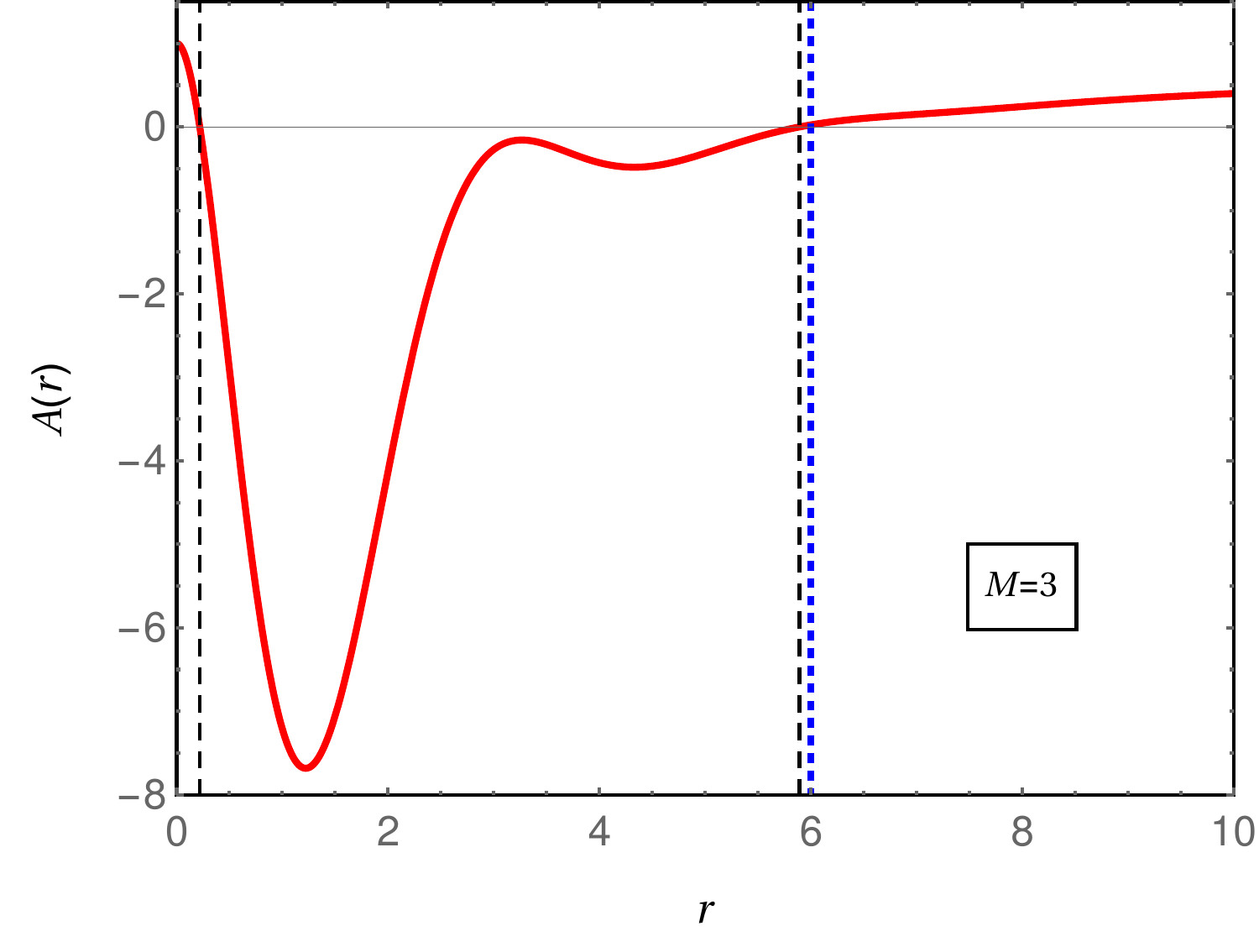}
\end{subfigure}%
\caption{\small Plots of the function $A(r)$ [Eq.~\eqref{Am}] in the case $a=1$ and $b=2$, corresponding to $q=2$, for the four possible scenarios described in the central rows of Table~\ref{Tab2}. All plots are in units $M_{\rm P}=1$; the dashed lines represent the horizon coordinates, while the blue dotted line represents the position of the Schwarzschild radius $r_\text{S}=2M$. In the first plot we have $M<0.345$, corresponding to a horizonless compact object. In the second one we have $0.34<M<2.25$, corresponding to a black hole with an inner horizon $r_{-}$ and an event horizon $r_\text{H} < r_\text{S}$. The third one, with $2.25<M<2.59$, represents an object with the maximal possible number of horizons, in this case $N_\text{H}=N_\text{H}^{\text{max}}=4$; the position of the event horizon is such that $r_\text{H} > r_\text{S}$. Finally, the last plot represents the case $M>2.59$, where we have a black hole with two horizons and 
$r_\text{H}$ approaches the Schwarzschild radius. }
\label{Fig3}
\end{figure}

As examples of the four regimes described above, in Table~\ref{Tab2} we display, for the particular cases of $q=1$, $q=2$ and $q=3$, the number of horizons for each range of values of $M$.
Each of these intervals can be viewed as a mass gap, beyond which the number of horizons changes. At the critical value between them, a pair of horizons merge into one extremal horizon.
Table~\ref{Tab2} offers an explicit verification of the results obtained so far, namely:
\begin{itemize}
\item[i.] For the case $q=1$, studied before in the literature~\cite{Bambi:2016wmo}, the sole possibilities are to have a horizonless object (if $M < 1.08 M_{\rm P}^2/a$) or a two-horizon black hole (if $M > 1.08 M_{\rm P}^2/a$).
\item[ii.] For $q=2$ the metric can have at most four horizons. However, since the effective mass function does not assume negative values for $q<2.67$ (see Sec.~\ref{Sec2.massfunc}), only the trivial pair of horizons remains if $M$ is too large, as shown also in Fig.~\ref{Fig3}.
\item[iii.] For $q=3$ the function $\tilde{m}(y)$ has one negative minima, which contributes a non-trivial pair of horizons in the limit of large $M$. In fact, for $M > 3.16 M_{\rm P}^2/a$ the metric has exactly four horizons, while for smaller values of $M$ it can have up to six horizons.
\end{itemize}

Another example of the diversity of horizon configurations is provided by Fig.~\ref{Fig3}, where we show the graphs of $A(r)$ in~\eq{Az} for each of the characteristic scenarios of the case $q=2$. In those graphs we also mark the position of the Schwarzschild radius, motivating a comparison of the size of the Lee--Wick black hole and the Schwarzschild one.

\subsubsection{Position of the horizons and size of the black hole}
\label{Sec.PHSBH}

The presence of vertical asymptotes to the curve~\eq{Gzero} on the first quadrant of the $yq$-plane (see Fig.~\ref{Fig2}) makes it possible to establish bounds on the positions of some horizons. Indeed, 
if $n_{\text{E}}>1$,
the last maximum of $Z_q(y)$ occurs somewhere in between $y=(2n_{\text{E}}-1)\pi/2$ and $y=n_{\text{E}}\pi$, while the other maxima tend to occur very close to odd multiples of $\pi$; see Appendix~\ref{Appendix} for the details. On the other hand, if $n_{\text{E}}=1$ the only maximum of $Z_q(y)$ occurs for $y \in (0,\pi)$.
Recalling that the maxima of $Z_q(y)$ form a decreasing sequence, this leads to the conclusion that the $(2k+1)$-th horizon (counting outwards) occurs for $y$ in the interval $(2k\pi, (2k+1)\pi]$. 
Using~\eq{y_e_q} to return to the coordinate $r$, it follows, for example, that the first horizon occurs for $r \in (0 , \pi/b ]$ and the third one, for $r \in (2\pi/b , 3\pi/b ]$. Hence, the radii of these horizons are inversely proportional to the parameter $b$. However, note that the innermost horizon can never reach the point $r = 0$ because Eq.~\eq{oslimites} implies that $A(0) = 1$.

In what concerns the even-numbered horizons, the position of any inner even horizon is bounded by the previous and next horizons. This observation has an important consequence for the position $r_\text{H}$ of the largest horizon, as it defines a sequence of \emph{position gaps} for the outermost horizon.
Indeed, around $r=0$ and each local maximum of $A(r)$ there exists an interval such that if an horizon lies within this interval, then it must be an inner horizon. Hence, the outermost horizon cannot fall in any of these intervals and black holes exist in different regimes of radius, separated by these gaps.

For a fixed $q$, the number of horizon position gaps is given by $N_\text{H}^{\text{max}}/2$, and one can use Eq.~\eq{HmaxAPP} for an estimation. To prove this result, notice that $N_\text{H}^{\text{max}}/2$ is the maximal number of odd-numbered horizons, which coincides with the number of local maxima of $A(r)$ plus one (because of the point $r=0$). Since the first gap is around $r=0$, the number of possible black hole regimes is equal to the number of position gaps. For example, for $q=2$ we have two gaps and two regimes of black holes; the second panel of Fig.~\ref{Fig3} corresponds to a ``small black hole'', while the last two panels represent ``large black holes''. 

The existence of horizon gaps is a general feature of metrics with an oscillating function $A(r)$ (see~\cite{Nicolini:2019pgi} for another example, in the context of higher-dimensional gravity). 
In the case of Lee--Wick black holes, this means that for $q>1.67$ there are multiple regimes for the size of black holes, separated by intermediate intervals of $r_\text{H}$ where black hole solutions are not allowed. 
In Sec.~\ref{Sec5} we shall elaborate more on the sequence of horizon position gaps and their consequence to the thermodynamics of Lee--Wick black holes.

We close this section by comparing the size of Lee--Wick and Schwarzschild black holes.
In general, owing to the oscillations of the effective mass function, the Lee--Wick black holes can be either smaller or larger than a Schwarzschild one. In fact, from Eq.~\eqref{Am} it is easy to see that if the outer (event) horizon occurs in a region where ${m}(r)<M$, then this horizon is smaller than the Schwarzschild radius $r_\text{S} = 2GM$; on the other hand, if ${m}(r)>M$
at the outer horizon, then $r_\text{H} > r_\text{S}$. 
Moreover, if the outer horizon occurs in a region where the oscillations are already damped, we have that ${m}(r)\approx M$ and there is no upper bound for the position $r_\text{H}$ of the largest horizon, which 
becomes roughly the same as the Schwarzschild radius. The last regime coincides with the case in which $M > M_\text{c}$, in other words, if the mass $M$ of the object is much larger than $M_{\rm P}^2 /a$ (assuming that $a$ and $b$ have similar orders of magnitude, see the discussion in Appendix~\ref{Appendix}).
All these possibilities can be viewed in Fig.~\ref{Fig3}, where we display the graphs of $A(r)$ for the case $q=2$ and different values of $M$.


\subsection{Regularity of the solution}
\label{Sec2.reg}

\begin{figure}[b]
\includegraphics[width=8cm]{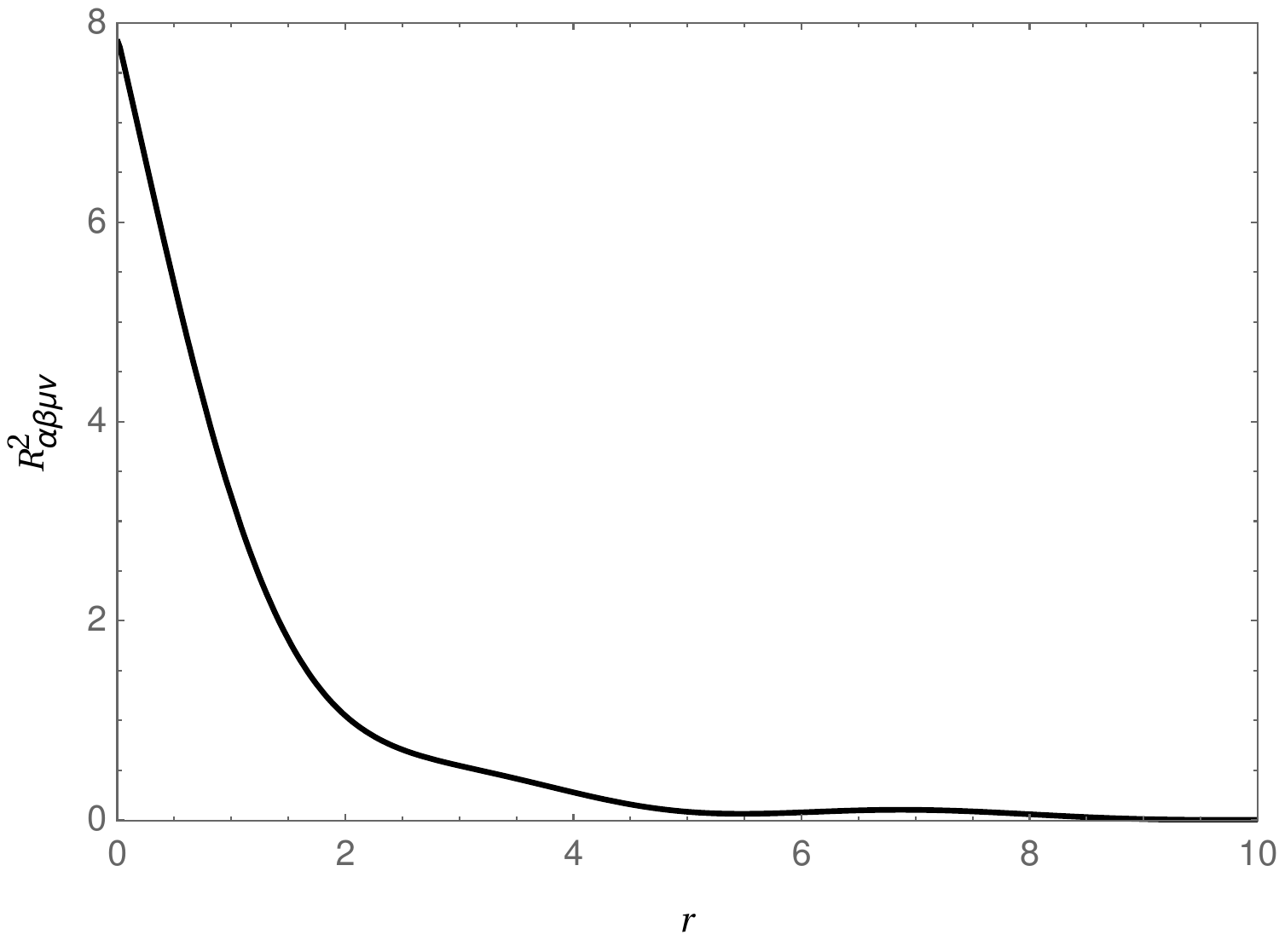}
\caption{\small Kretschmann scalar for the case $a=0.25$, $b=0.77$ (\textit{i.e.}, $q=3.08$), in units with $G=M=1$.} \label{KretsP}
\end{figure}

In {either case} the solution~\eq{metricB=0} with~\eq{Am} describes a black hole or a horizonless object, it is regular in the sense that the curvature invariants without covariant derivatives are bounded everywhere. 
This can be proved by noticing that all the components of ${R^{\mu\nu}}_{\al\be}$ are finite --- see Appendix~\ref{AppendixB} for the explicit expressions --- so, the invariants constructed from the Riemann and the metric tensors are bounded~\cite{Bronnikov:2012wsj}. For example, expanding the Kretschmann scalar around $r=0$ we get
\beq
\label{Kre1}
R_{\mu\nu\al\be}^2 = \frac{8 (a^2+b^2)^4 (GM)^2}{3 a^2} + O(r) 
\eeq
(a plot of $R_{\mu\nu\al\be}^2$ is shown in Fig.~\ref{KretsP}). 
The finiteness of the curvature invariants at $r=0$ can also be explained by the small-$r$ behavior of the effective mass~\eq{mrsmall}, as it causes
the term linear on $r$ to be absent from the Taylor series of $A(r)$ around $r=0$ 
(see, \textit{e.g.},~\cite{Bronnikov:2012wsj,Frolov:2016pav,Giacchini:2021pmr}).
In other words, the metric has a de Sitter core.

On the other hand, curvature invariants \emph{with} covariant derivatives can be singular. This is owing to the presence of the term cubic in $r$ in the Taylor series of the function $A(r)$~\cite{Giacchini:2021pmr}. For instance, the invariant
\beq
\label{BoxK1}
R_{\mu\nu\al\be} \Box R^{\mu\nu\al\be} = -\frac{(a^2+b^2)^4 (GM)^2}{a} \left( \frac{20}{3 r} - \frac{19 a^2-4 b^2 }{a} \right)  + O(r) 
\eeq
diverges as $r\to 0$.


\section{Dirty Lee--Wick black holes}
\label{Sec3}

It is possible to obtain families of static spherically symmetric Lee--Wick black holes different from~\eq{metricB=0}  by assuming that the metric has a nontrivial shift function $B(r)$, namely,
\beq
\label{metricB}
\rd s^2 = - A(r) e^{B(r)} \rd t^2 + \frac{\rd r^2}{A(r)} + r^2 \rd \Omega^2 .
\eeq
We shall refer to this solution as ``dirty'' because the function $B(r)$ is nontrivial, similarly to what happens in the so-called ``dirty black holes'' which are in interaction with matter fields~\cite{Visser:1992qh}.
In this case, instead of~\eq{System}, the effective field equations are
\begin{subequations} 
\label{SystemB}
\begin{align} 
\label{primeira}
& \frac{1}{r} \frac{\rd A}{\rd r} +\frac{A-1}{r^2} = - 8\pi G \rho
,
\\
&
\frac{A}{r} \frac{\rd B}{\rd r} = 8\pi G ( \rho + p_r )
,
\\
&
\frac{\rd p_r}{\rd r} = \frac{2}{r} \left( p_\th - p_r \right) - \frac{\left( p_r + \rho \right) }{2} \left( \frac{1}{A}\frac{\rd A}{\rd r} + \frac{\rd B}{\rd r} \right) .
\label{ultima}
\end{align}
\end{subequations}

This system must be supplemented by an equation of state for the effective pressures; in Ref.~\cite{NosG} several possibilities have been discussed in the general framework of higher-derivative gravity models.\footnote{See also~\cite{dirty,Mureika:2010je,Gaete:2010sp} for similar procedures in other frameworks.} 
The solution we present here follows from the equation of state
\beq
\label{EOS}
p_r(r) = \left[  A(r) - 1 \right] \rho(r) ,
\eeq
which was first considered in~\cite{BreTibLiv} and studied in more detail in~\cite{NosG}.
For the effective source~\eq{effsource}, it results in the solution for $B(r)$,
\beq
\begin{split}
\n{Bsol}
B(r) & =  8 \pi G \int_\infty^r \rd x \, x \, \rho (x) 
\\
& =  -\frac{G M (a^2+b^2) }{a b} e^{-a r} \left[ b \cos(b r) + a \sin (b r)\right] 
.
\end{split}
\eeq
On the other hand, the solution of Eq.~\eq{primeira} for $A(r)$ is the same as in Sec.~\ref{Sec2}, given by~\eq{Am} with the mass function $m(r)$ of Eq.~\eq{meff}; while the tangential pressure $p_\th$ can be determined from Eq.~\eqref{ultima}. 
Therefore, the final solution reads
\beq
\label{metricBfinal}
\rd s^2 = - \left(1-\frac{2G m(r)}{r} \right) \exp \left\lbrace    -\frac{G M (a^2+b^2) }{a b} e^{-a r} \left[ b \cos(b r) + a \sin (b r)\right]   \right\rbrace  \, \rd t^2 
+ \frac{\rd r^2 }{{\left(1-\frac{2G m(r)}{r} \right)}} 
+ r^2 \rd \Omega^2
.
\eeq

The metric~\eq{metricBfinal} has exactly the same structure of horizons and mass gaps as the solution considered in Sec.~\ref{Sec2}, since in both cases the horizons are defined by the same equation, namely,  $A(r)=0$. 

An important difference between the two families of Lee--Wick black holes is regarding the weak gravitational field regime. 
First, note that for a small mass $M$, \textit{i.e.}, if we drop the terms $O(M^2)$, the metric~\eq{metricBfinal} boils down to 
\beq
\label{metricBlin}
\rd s^2 = - \left( 1 + 2\ph \right) \rd t^2 
+ \left( 1 + 2\ph^\prime r \right) \rd r^2
+ r^2 \rd \Omega^2 
,
\eeq
where 
\beq
\n{mo-pot}
\ph(r) = -\frac{G M}{r} \left[ 1 - \frac{2 a b \cos(b r) + (a^2-b^2) \sin (b r)}{2 a b} e^{-a r} \right] 
\eeq
is precisely the modified Newtonian potential for the model~\eq{action}, evaluated in~\cite{Accioly:2016qeb}.\footnote{Indeed, for the higher-derivative model~\eq{action}, performing the expansion $g_{\mu\nu} = \eta_{\mu\nu} + h_{\mu\nu}$, it is possible to show that, at first order in $h_{\mu\nu}$, a static and spherically symmetric metric can always be expressed in terms of only one function $\ph(r)$. For a point-like source it is the solution of the modified Poisson equation
$$
f( \De) \De \ph (r) = 4\pi G \delta (\vec {r})
$$   
or, equivalently (by inverting the operator $f(\Delta)$),
$$
\De \ph (r) = 4\pi G \rho (r),
$$
where $\rho (r)$ is the effective source~\eq{effsource}; see, \textit{e.g.},~\cite{BreTib2,BreTibLiv}.} This does not happen with the solution in the form~\eq{metricB=0} of the previous section, although both metrics tend to the linearised Schwarzschild metric in the far infrared limit. 
A general discussion about this issue in the context of higher-derivative gravity can be found in~\cite{NosG}, where it was proved that, differently from the case $B(r) \equiv 0$, the solution of~\eq{SystemB} associated with the equation of state~\eq{EOS} matches the modified Newtonian-limit metric sourced by a point-like particle in the linear regime.

Also, with respect to the small-$r$ behaviour, it can be shown that the metric~\eq{metricBfinal} has the same \emph{qualitative} behaviour as the solution~\eq{metricB=0} of the previous section, \textit{i.e.}, all the curvature invariants without covariant derivatives are finite, while the ones with covariant derivatives might diverge. Nevertheless, these invariants are quantitatively different for each solution. For example, around $r=0$ we get
\beq
R_{\mu\nu\al\be}^2 = \frac{5 (a^2+b^2)^4 (GM)^2 }{3 a^2}   + O(r) ,
\eeq
and
\beq
R_{\mu\nu\al\be} \Box R^{\mu\nu\al\be} = 
- \frac{(a^2+b^2)^4 G^2 M^2}{3 a} \left[ \frac{10}{r} - \frac{24 a^3-5 a b^2 - 9 \left(a^2+b^2\right)^2 G M}{a^2} \right]  + O(r) ,
\eeq
which display the same behaviour of Eqs.~\eq{Kre1} and~\eq{BoxK1} but with different coefficients.

For the sake of completeness, the components of ${R^{\mu\nu}}_{\al\be}$ for the metric~\eq{metricBfinal} are explicitly calculated in Appendix~\ref{AppendixC}, where we show that they are bounded, guaranteeing the regularity of all invariants polynomial in curvatures. 
It is worthwhile to notice that these components are linear in $GM$ at $r=0$. Together with the fact that the metric~\eq{metricBfinal} coincides with the Newtonian-limit metric~\eq{metricBlin} to the leading order in $GM$,
this leads to the conclusion that the curvature invariants for the solution~\eq{metricBfinal} have the same small-$r$ behaviour as those calculated with the modified Newtonian solution~\eq{metricBlin}.

\begin{figure}[b]
\includegraphics[width=8cm]{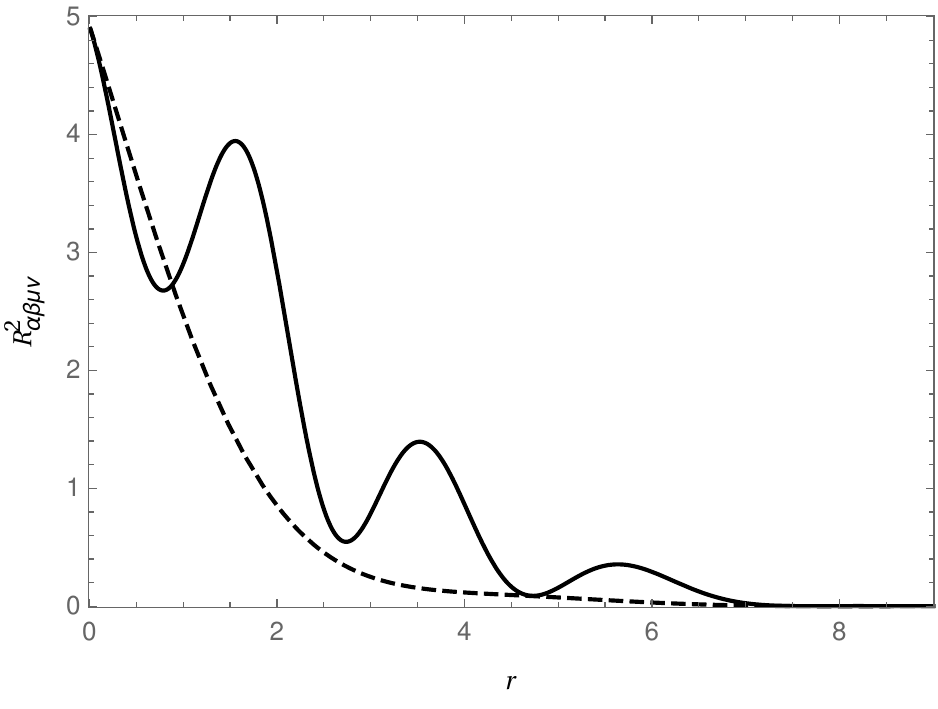}
\caption{\small 
Kretschmann scalar evaluated with the metric~\eq{metricBfinal} (solid line) and the linearised Kretschmann associated with~\eq{metricBlin} at leading order in $M$ (dashed line), for the case $a=0.25$, $b=0.77$ (\textit{i.e.}, $q=3.08$), in units with $G=M=1$. As explained in the text, in the regimes of small and large $r$ the behaviours of the two solutions are similar.
\label{KretsPB}
}
\end{figure}

Hence, one can say that the solution presented in this section matches the modified Newtonian-limit solution in both the regimes of large and small $r$.
This can be viewed, \textit{e.g.}, in Fig.~\ref{KretsPB}, which compares the Kretschmann scalar for the linearised and the non-linearised solutions. In addition, the comparison with Fig.~\ref{KretsP} shows that the solution~\eq{metricB=0} of the previous section does not have this property.


\section{Black hole thermodynamics}
\label{Sec5}

The oscillations of the function $Z_q(y)$~\eqref{Z} have interesting consequences to the thermodynamic properties of the black hole solutions of Secs.~\ref{Sec2} and \ref{Sec3}. In particular, it causes the Hawking temperature to oscillate, leading to a varied phenomenology in which multiple mass regimes for black hole remnants are allowed, as we show in this section.

Before introducing the Hawking temperature for our system we recall that, once $a$ and $q$ are specified, the value of $M$ determines the number and the position of the horizons. Hence, if $r_\text{H}$ is the position of the  outermost horizon, 
the  equation $A(r_\text{H})=0$ can be solved for $M$ [see Eq.~\eq{Az}], resulting in a relation between the mass of the black hole and the position of the horizon, namely,
\beq \label{HorCondM}
M(y_\text{H})=\frac{M_{\rm P}^2}{a \, Z_{q} (y_\text{H})},
\eeq 
where $y_\text{H} \equiv a q r_\text{H}$, as usual.

\begin{figure}[t]
\centering
\begin{subfigure}{.33\textwidth}
\centering
\includegraphics[width=5.3cm]{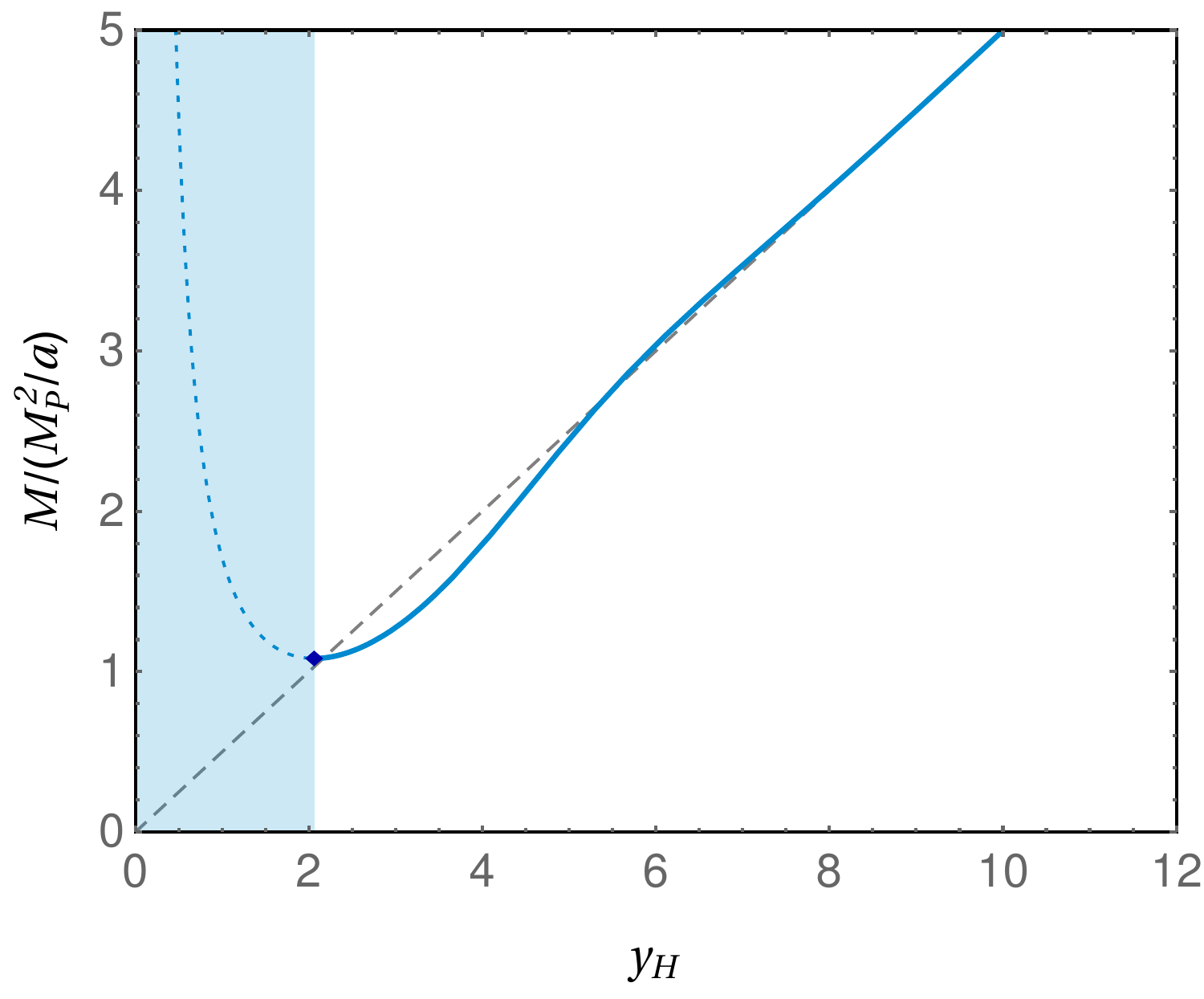}
\caption{\small $M(y_\text{H})$ for $q=1$.}
\label{Fig8a}
\end{subfigure}%
\begin{subfigure}{.33\textwidth}
\centering
\includegraphics[width=5.3cm]{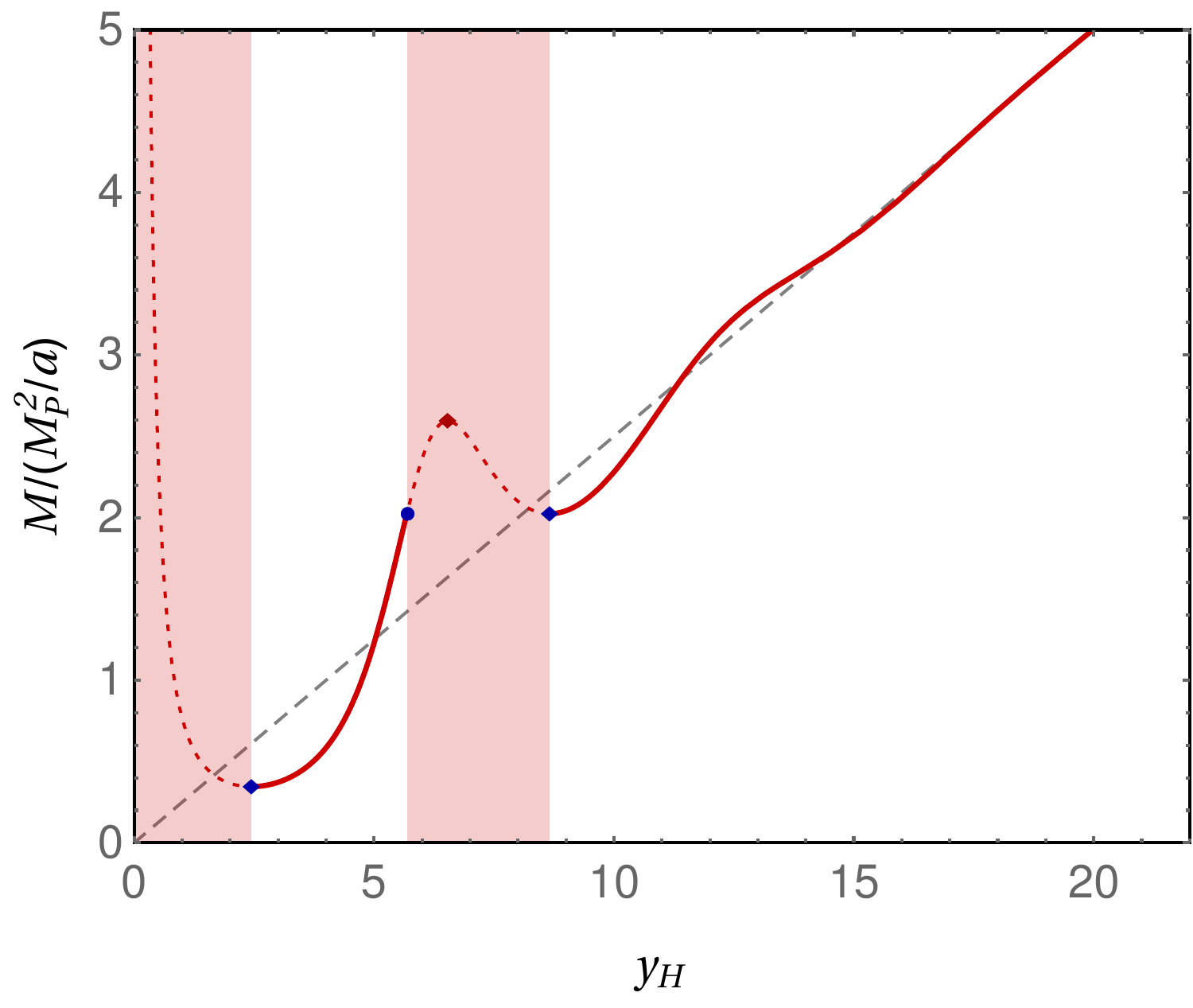}
\caption{\small $M(y_\text{H})$ for $q=2$.}
\label{Fig8b}
\end{subfigure}
\begin{subfigure}{.33\textwidth}
\centering
\includegraphics[width=5.3cm]{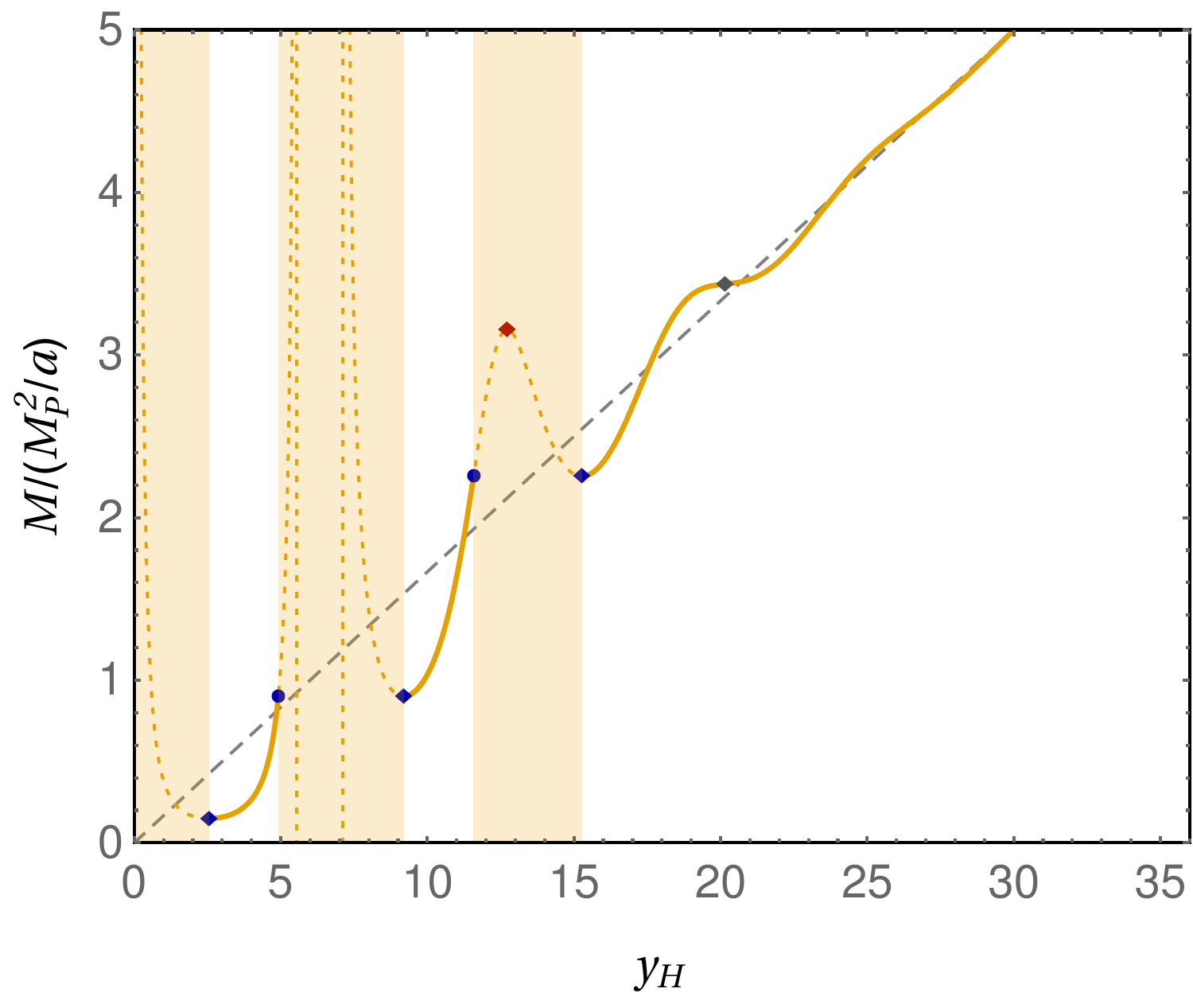}
\caption{\small $M(y_\text{H})$ for $q=3$.}
\label{Fig8c}
\end{subfigure}
\\
\vspace{0.5cm}
\begin{subfigure}{.33\textwidth}
\centering
\includegraphics[width=5.3cm]{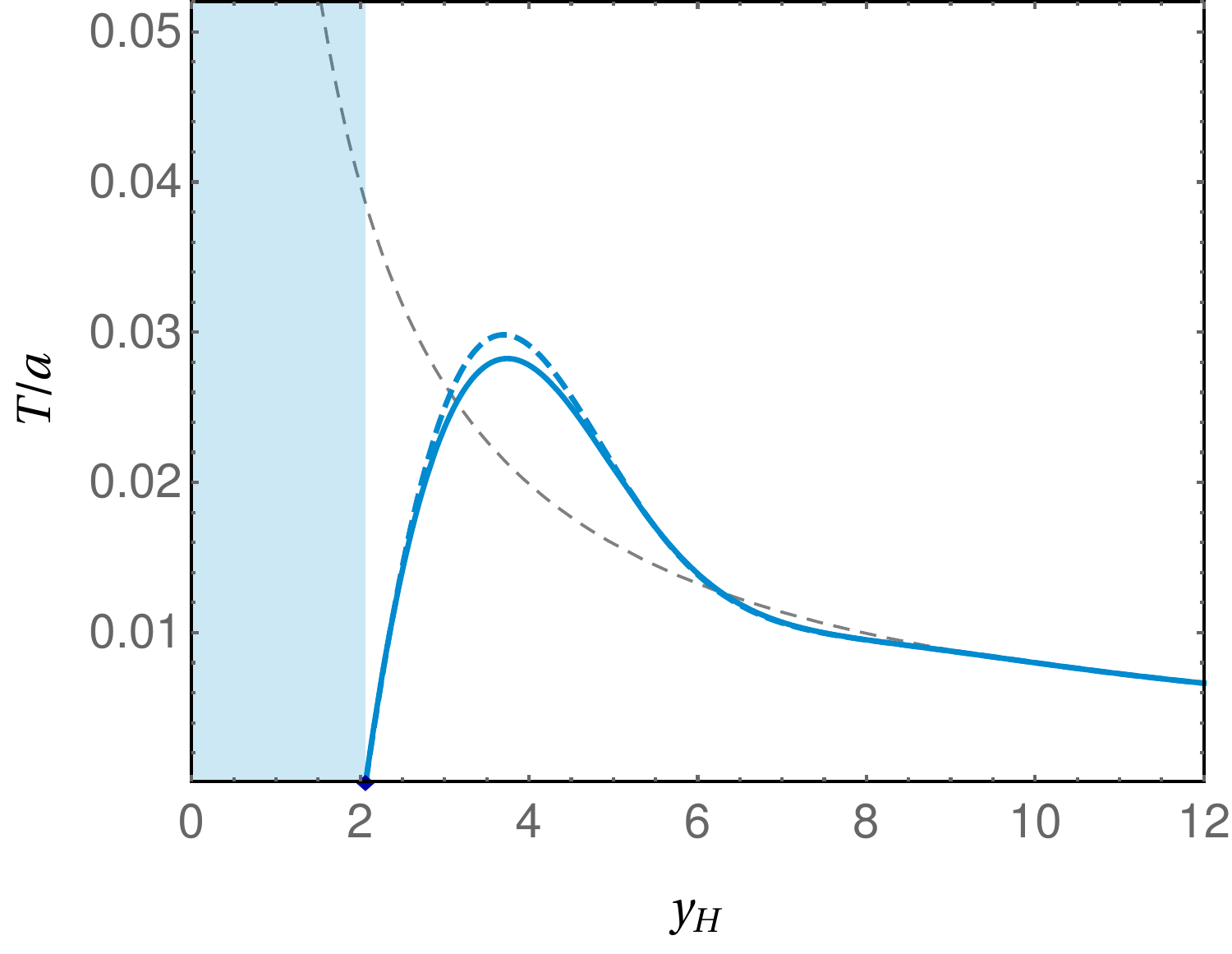}
\caption{\small $T(y_\text{H})$ for $q=1$.}
\label{Fig8d}
\end{subfigure}%
\begin{subfigure}{.33\textwidth}
\centering
\includegraphics[width=5.3cm]{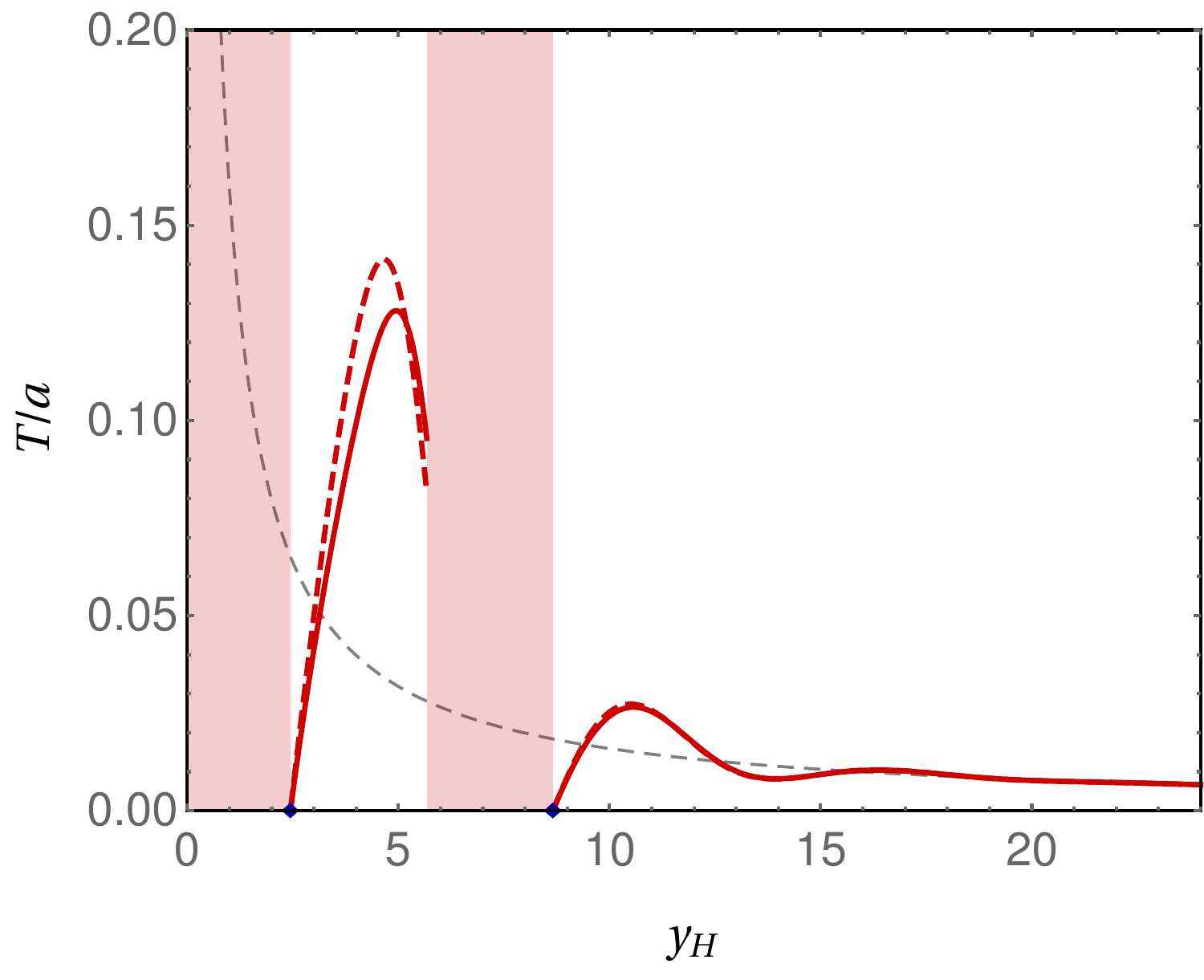}
\caption{\small $T(y_\text{H})$ for $q=2$.}
\label{Fig8e}
\end{subfigure}
\begin{subfigure}{.33\textwidth}
\centering
\includegraphics[width=5.3cm]{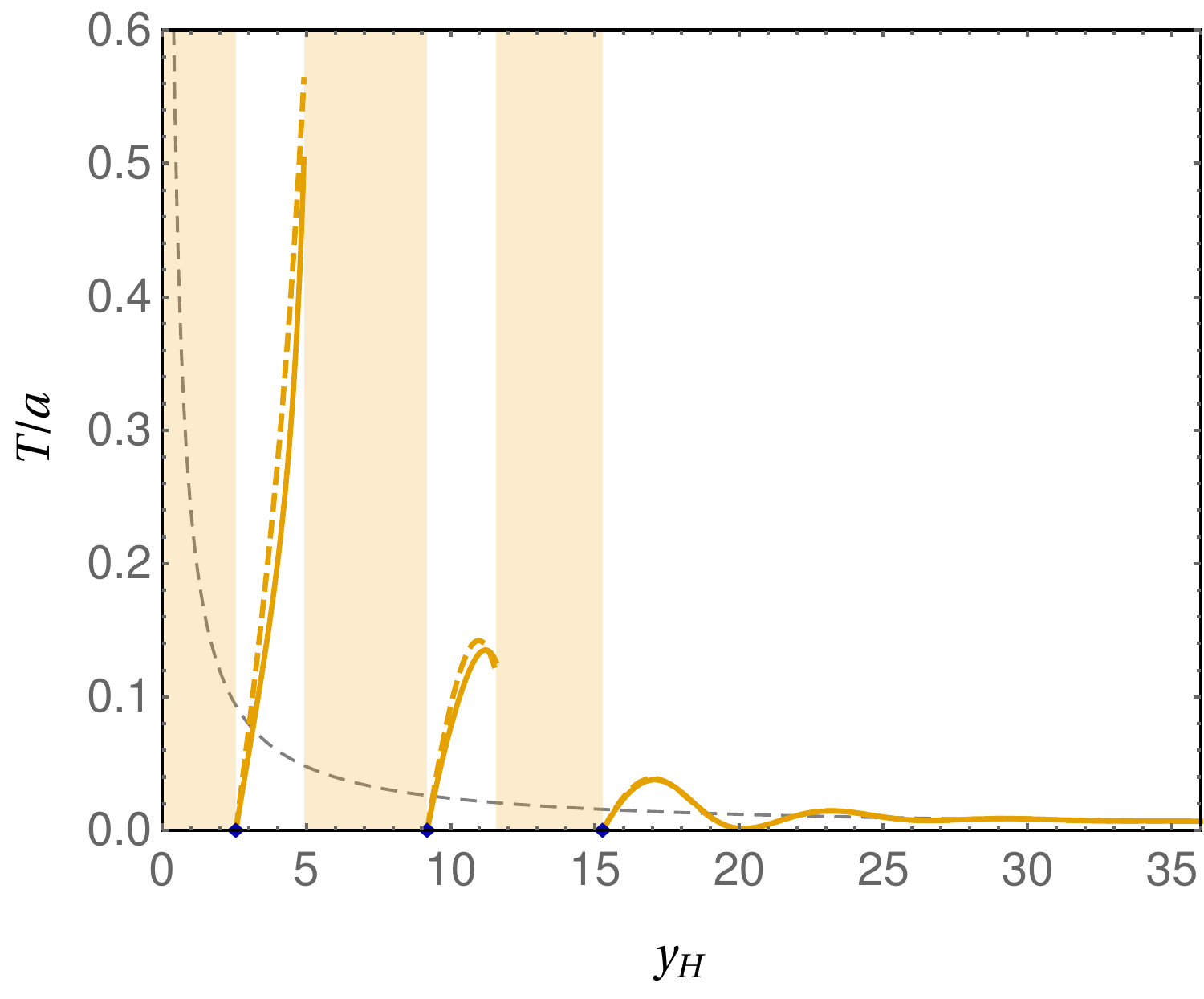}
\caption{\small $T(y_\text{H})$ for $q=3$.}
\label{Fig8f}
\end{subfigure}
\\
\vspace{0.5cm}
\begin{subfigure}{.33\textwidth}
\centering
\includegraphics[width=5.3cm]{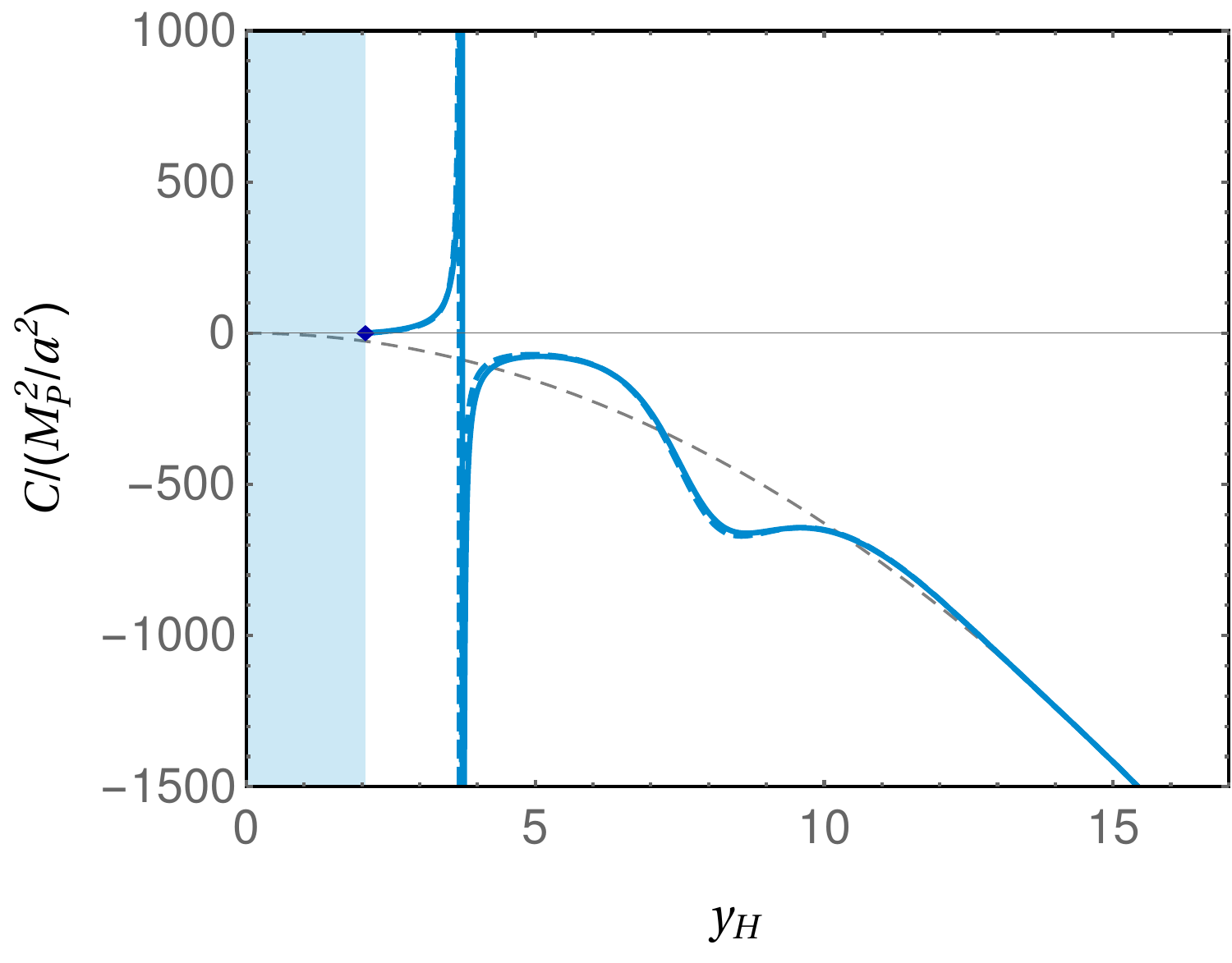}
\caption{\small $C(y_\text{H})$ for $q=1$.}
\label{Fig8g}
\end{subfigure}%
\begin{subfigure}{.33\textwidth}
\centering
\includegraphics[width=5.3cm]{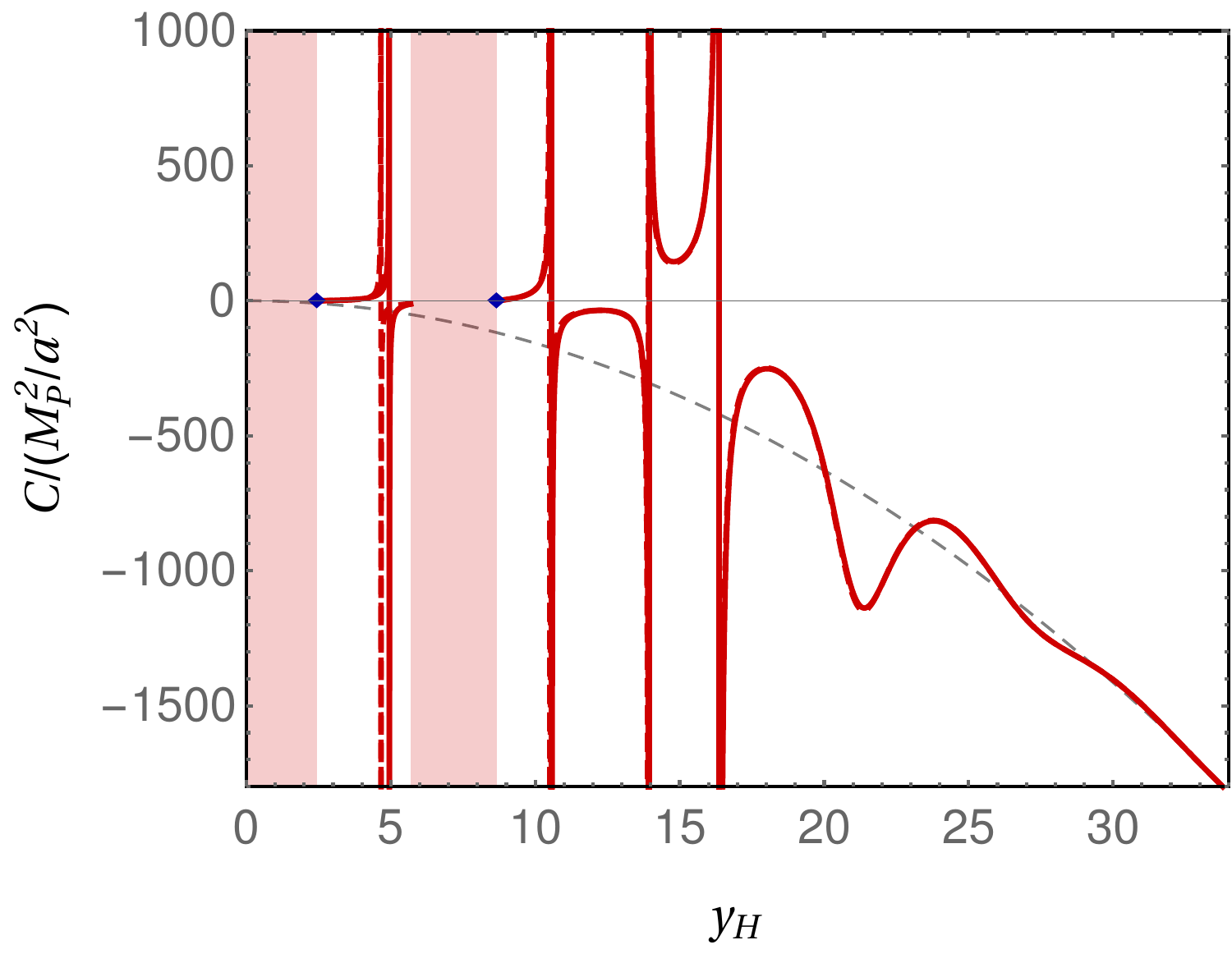}
\caption{\small $C(y_\text{H})$ for $q=2$.}
\label{Fig8h}
\end{subfigure}
\begin{subfigure}{.33\textwidth}
\centering
\includegraphics[width=5.3cm]{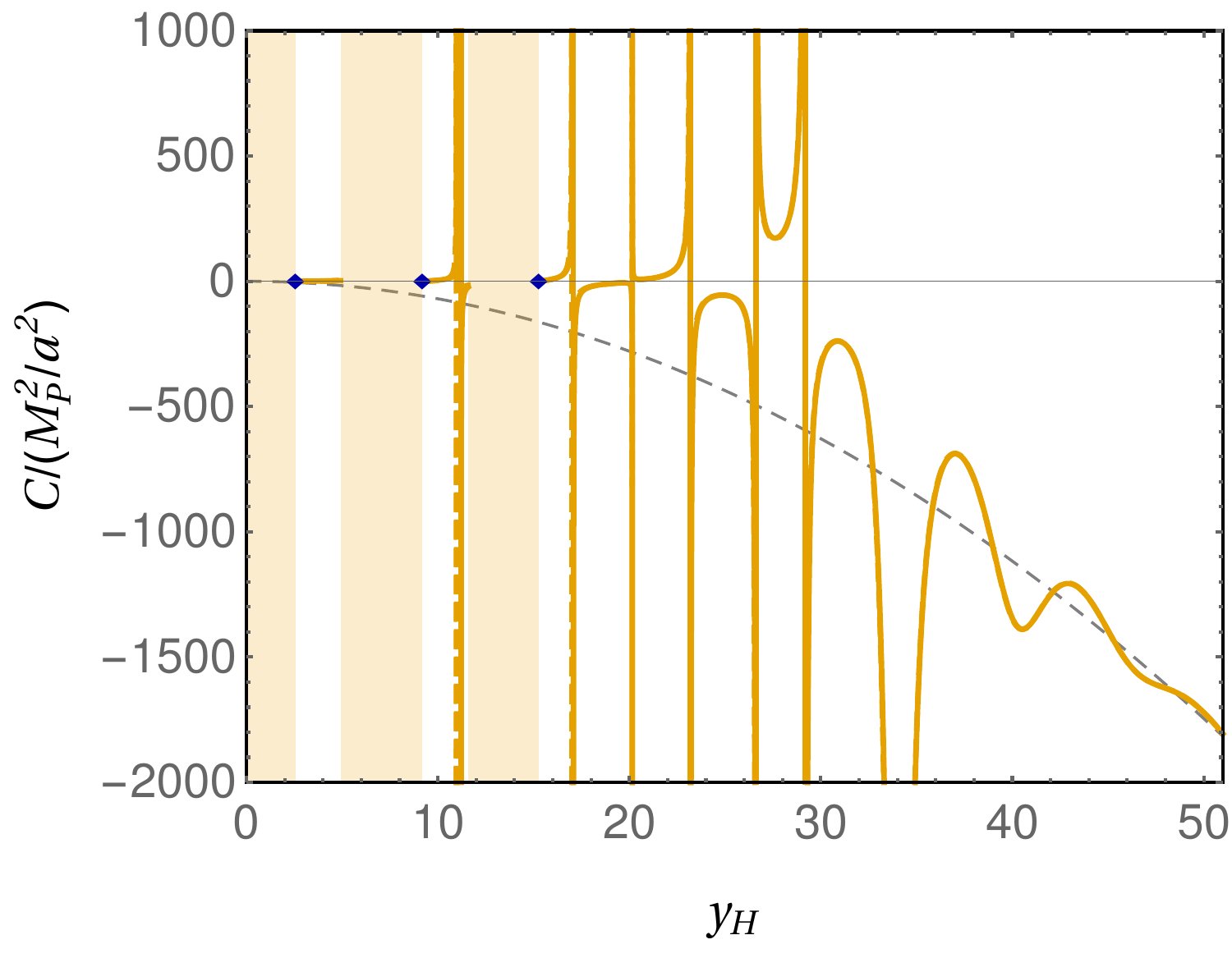}
\caption{\small $C(y_\text{H})$ for $q=3$.}
\label{Fig8i}
\end{subfigure}
\caption{\small 
Plots of the mass $M(y_\text{H})$~\eqref{HorCondM} (first row), Hawking temperature $T(y_\text{H})$~\eqref{hawkingTB} (second row), and heat capacity $C(y_\text{H})$~\eqref{heatCLW} (third row), as function of the horizon coordinate $y_\text{H}$, for $q=1$ (blue), $q=2$ (red) and $q=3$ (yellow). The mass is expressed in units of $M_{\rm P}^2/a$, the temperature in units of $a$ and the heat capacity in units of $M_{\rm P}^2/a^2$. The shaded regions where the functions are not defined represent the position gaps for the outer horizon. The thin dotted line in the panels (a), (b) and (c) fall in the forbidden regions and represent internal horizons; for the explanation of the circle and diamond markers in these panels see the captions of Table~\ref{Tab4} and  Fig.~\ref{Fig9}. The thick dashed lines in the graphs of $T(y_\text{H})$ and $C(y_\text{H})$ correspond to the dirty black holes of Sec.~\ref{Sec3}, while the solid lines refer to those of Sec.~\ref{Sec2}. In all the graphs, the thin dashed grey line represents the behaviour of the thermodynamic quantity for the Schwarzschild black hole.
}
\label{Fig8}
\end{figure}

In the first row of Fig.~\ref{Fig8} we show the plot of \eqref{HorCondM} in units of $M_{\rm P}^2/a$ for three different values of $q$, namely, $q=1,2,3$. We observe that the function $M(y_\text{H})$ (solid line) displays an increasing number of discontinuities for larger values of $q$, represented by the shaded areas in the graphs where $M(y_\text{H})$ is not defined. These regions are excluded because, even though there exists an $y$ that formally solves~\eqref{HorCondM}, it corresponds to an inner horizon, 
\textit{i.e.}, $y \neq y_\text{H}$ (the position of such inner horizons are represented by thin dotted lines in Figs.~\ref{Fig8a}-\ref{Fig8c}).

In other words, the excluded regions are the horizon position gaps described in Sec.~\ref{Sec.PHSBH}. Remember that
the occurrence of inner horizons coincides with the presence of intervals of radius where the black hole is not physically realized. In the case of regular black holes with a single mass gap $M_0$ (see Sec.~\ref{Sec.ANH}), this statement trivially means that there is a minimum radius $r_0$ that equals the radius where a pair of horizons merge into one extremal horizon, and no black hole can exist with $r_\text{H} < r_0$. Similarly, if the function $M(y_\text{H})$ has multiple local minima, each one corresponding to a critical mass $M_{0}, M_{1}, \ldots$, there exist intervals of values of $r_\text{H}$ that do not correspond to any physical black hole. For example, in the case $q=2$, no black hole can have a radius $5.68 <y_\text{H}<8.66 $, and two objects with masses slightly smaller and  slightly larger than $M = 2.02 M_{\rm P}^2/a$ would have event horizons with very different radii.

\begin{table}[b]
    \centering
    \begin{tabular}{|c|c|c|c|}
            \hline
                & Critical masses & \multirow{2}{*}{Extremal horizon}  & \multirow{2}{*}{Black hole regimes} \\ 
                & (in units of $M_{\rm P}^2/a$) &   & \\
            \hline
            \hline
            \multirow{1}{*}{$q=1$} & \multicolumn{1}{c|}{$M_0=1.082$} & \multicolumn{1}{c|}{$y_0=2.02$} & \multicolumn{1}{c|}{Single regime: $M > M_0$ and $y_\text{H} > y_0$}  \\\cline{2-4}
            \hline
            \multirow{2}{*}{$q=2$} & \multicolumn{1}{c|}{$M_0=0.345 $} & \multicolumn{1}{c|}{$y_0=2.44$} & \multicolumn{1}{c|}{Small black hole: $M_0 < M < M_1$ and $2.44 < y_\text{H} < 5.68$} \\\cline{2-4}
                                 & \multicolumn{1}{c|}{$M_1=2.024 $} & \multicolumn{1}{c|}{$y_1=8.66$} & \multicolumn{1}{c|}{Large black hole: $M > M_1$ and $y_\text{H} > 8.66$} \\\hline                                 
            \hline
            \multirow{3}{*}{$q=3$} & \multicolumn{1}{c|}{$M_0=0.149 $} & \multicolumn{1}{c|}{$y_0=2.55$} & \multicolumn{1}{c|}{Small black hole: $M_0 < M < M_1$ and $2.55 < y_\text{H} < 4.92$} \\\cline{2-4}
                                 & \multicolumn{1}{c|}{$M_1=0.901 $} & \multicolumn{1}{c|}{$y_1=9.18$} & \multicolumn{1}{c|}{Intermediate black hole: $M_1 < M < M_2$ and $9.18 < y_\text{H} < 11.58$}\\\cline{2-4}
                                 & \multicolumn{1}{c|}{$M_2=2.254 $} & \multicolumn{1}{c|}{$y_2=15.27$} & \multicolumn{1}{c|}{Large black hole: $M > M_2$ and $y_\text{H} > 15.27$} \\\hline                                
        \end{tabular}    
        \caption{\small Values of critical masses of extremal black holes, radii of the extremal outer horizon and possible black hole regimes for $q=1,2,3$. 
        The pairs $(y_i, M_i)$ in this table correspond to the blue diamonds in Figs.~\ref{Fig8a}-\ref{Fig8c} and they mark the beginning of the allowed range for $y_\text{H}$, whereas the upper bound of the allowed regions are marked by blue circles in Figs.~\ref{Fig8a}-\ref{Fig8c}.
        Extremal inner horizons are not listed in this table as they do not affect the position gaps for the outermost horizon, but in Figs.~\ref{Fig8b} and~\ref{Fig8c} they are represented by red diamonds.
         }
\label{Tab4}
\end{table}

As another example, in Table~\ref{Tab4} we display the critical masses, extremal horizons and the possible regimes for black hole masses and radii for $q=1,2,3$. Since the number of black hole regimes is $N_\text{H}^{\text{max}}/2$, from Eq.~\eq{HmaxAPP} we expect to have, respectively, one, two and three possible regimes --- which is confirmed in the last column of the table. It is also instructive to compare the Tables~\ref{Tab2} and~\ref{Tab4}: notice that the black hole regimes are defined by the smaller critical masses of Table~\ref{Tab2}. On the other hand, the higher critical masses corresponding to the decrease of the number of horizons are always associated to \textit{inner} extremal horizons and they do not impose any restriction on the position of the outermost horizon; for this reason, they do not define new regimes in Table~\ref{Tab4} and such extremal inner horizons were omitted there.

Having identified the possible regimes for a Lee--Wick black hole, we can consider the evaporation process.
To this end, let us determine the surface gravity $\kappa$. 
For a generic static spherically symmetric metric in the form~\eqref{metricB} we have
\bea
\kappa & = & \frac{1}{2} \frac{1}{\sqrt{e^{B(r)}}}\frac{\rd}{\rd r} \left[ A(r) e^{B(r)} \right] \bigg\vert_{r=r_\text{H}} 
\nonumber
\\
& = & \cfrac{1}{2} \,e^{\frac{1}{2}B (r_\text{H})} \,  A'(r_\text{H})   \;, 
\label{surgrav}
\eea
where we used 
$A(r_\text{H})=0$. This formula is valid not only for the dirty Lee--Wick black holes of Sec.~\ref{Sec3}, but also to the solutions of Sec.~\ref{Sec2}; in the latter case, it suffices to set $B(r)\equiv 0$.

The Hawking temperature of the black hole,
\beq
\n{T_au}
T = \frac{\ka}{2\pi},
\eeq 
can then be obtained from~\eq{surgrav}. Taking into account formulas~\eq{Az} and~\eq{HorCondM} it follows
\beq \label{hawkingTB}
T(y_\text{H})=-\frac{a q}{4 \pi} \frac{Z_q'(y_\text{H}) }{ Z_q(y_\text{H})} \, e^{\frac12 B_q(y_\text{H})} .
\eeq
The function $B_q(y_\text{H})$ is non-trivial only in the case of dirty black holes, in which it
is the analogous of $B(r)$ in Eq.~\eq{Bsol} written in term of the dimensionless horizon coordinate $y_\text{H} = a q r_\text{H}$, namely,
\beq\label{B}
B_q(y_\text{H})=-\frac{1}{Z_q(y_\text{H})}\frac{ 1+q^2}{q} e^{-\frac{y_\text{H}}{q}} (q \cos y_\text{H}+ \sin y_\text{H} ) .
\eeq

Finally, for the heat capacity $C(y_\text{H})$ of the black hole we obtain
\beq
C = \frac{\partial M}{\partial T} 
= - 8 \pi \frac{M_{\rm P}^2}{a^2 q} \left( \frac{Z_q'}{ 2 Z_q'^2 - 2 Z_q Z_q'' - Z_q Z_q' B_q'} \right)\, e^{-\frac{B_q}{2}} .
\label{heatCLW}
\eeq
Of course, like the function $M(y_\text{H})$, the temperature and heat capacity of the black holes are not defined in the gap regions mentioned above, where $y_\text{H}$ does not correspond to a black hole solution. This can be seen as the shaded domains in the second and third rows of Fig.~\ref{Fig8}, which display, respectively, the functions $T(y_\text{H})$ and $C(y_\text{H})$ for $q=1,2,3$. Since these quantities depend on the shift function, the solid lines represent the black hole solution with $B(r)\equiv 0$ (discussed in Sec.~\ref{Sec2}), whereas the dashed lines correspond to the dirty Lee--Wick black hole of Sec.~\ref{Sec3}. These plots suggest that both types of solution have a very similar qualitative behaviour in what regards thermodynamics. Again, notice that for larger values of $q$ the oscillations of the function $Z_q(y)$ produce stronger oscillations also in the thermodynamic quantities.

In the previous sections we showed that the Lee--Wick black holes approach the Schwarzschild solution in the regions of large enough $r$. Moreover, if $M$ is much larger than the massive parameters of the model, $r_\text{H}$ tends to the Schwarzschild radius. Therefore, for $r_\text{H}$ sufficiently large we expect that the quantities in Eqs.~\eq{HorCondM}, \eq{surgrav}, \eq{hawkingTB} and~\eq{heatCLW} recover the results for the Schwarzschild black hole, namely,
\beq
\label{MTC_Sch}
M_\text{S}(r_\text{H}) = \frac{M_{\rm P}^2}{2} r_\text{H} , \qquad T_\text{S}(r_\text{H}) = \frac{1}{4\pi r_\text{H}} , \qquad C_\text{S}(r_\text{H}) = -2 \pi M_{\rm P}^2 r_\text{H}^2 .
\eeq
This indeed happens, as it can be proved by noticing that, for large arguments, $Z_q(y_\text{H}) \approx 2q/y_\text{H} = 2/(a r_\text{H}) $ and $B(r_\text{H}) \to 0$, and substituting these results in the previous expressions. It can also be verified in Fig.~\ref{Fig8}, where we display the behaviour~\eqref{MTC_Sch} for the Schwarzschild black hole as a thin dashed grey line.

The relation between the function $M(y_\text{H})$ and the Hawking temperature can be obtained by combining
Eqs.~\eqref{HorCondM} and~\eqref{hawkingTB}, which yields 
\beq \label{first_Principle}
\frac{\rd M}{\rd y_\text{H}} = \frac{4\pi}{a q} \, M(y_\text{H}) \, T(y_\text{H}) \, e^{- \frac12 B_q(y_\text{H})}.
\eeq
Hence, since $B_q(y_\text{H})$ is a bounded function and $M(y_\text{H})>0$ for physical solutions, the quantity $\rd M/\rd y_\text{H}$ (possibly considered as a one-sided limit) can be zero if and only if $T=0$. From this we conclude that the minima of the function $M(y_\text{H})$, which represent critical masses, always correspond to stable zero-temperature configurations.
Therefore, in each of the allowed horizon position gaps, the black hole will undergo a process of evaporation emitting radiation and its mass will asymptotically reach the critical mass of that gap.

To estimate the evaporation time we can use the Stefan--Boltzmann law,
\beq\label{stefan}
L = \sigma \, \mathcal{A} \, T^4 ,
\eeq 
that relates the 
luminosity $L$ of the black hole with its temperature $T$ and surface area $\mathcal{A} = 4\pi r_\text{H}^2 $ of the outermost horizon. The factor $\si$ depends on the quantum details of the system
and can be neglected since our main goal is to probe the effect of the oscillations of the solution to the process of evaporation. Therefore, considering that all the luminosity comes from the loss of mass, we have 
\beq\label{stefan2}
\frac{\rd M}{\rd t } \sim  - 4\pi r_\text{H}^2 \, T^4 .
\eeq

It is instructive to recall that for the Schwarzschild black hole, $T\sim r_\text{H}^{-1} \sim M^{-1}$, so we have $\rd M / \rd t \sim - M^{-2}$. This differential equation can be promptly integrated from an initial time $t_i=0$ to a final time $t_f$ assuming 
 $M(t_i)=M_{i}$ and $M(t_f)=0$,  and it yields $t_f \sim M_{i}^3$, \textit{i.e.}, the black hole radiates all its mass in a finite amount of time. 
For the Lee--Wick black holes, on the other hand, the estimate for the lifetime $\Delta t \equiv t_f - t_i$ calculated from Eq.~\eqref{stefan2} is
\beq \label{BHlifetime}
\Delta t \sim - (4\pi)^3 \, \frac{M_{\rm P}^2}{a^4 q^3} \int_{r_i}^{r_c} \rd r' \frac{Z_q^2(aq r')}{r'^2 Z'^3_q(aq r')}\,e^{-2 B(r')}  ,
\eeq
where the integral is performed from the initial radius $r_i$ of the black hole up to the critical radius $r_c$, where the temperature reaches $T=0$ and the evaporation stops. The value of this critical radius depends on which gap the initial black hole is, and it corresponds to one of the extremal horizons, as explained above. Therefore, we expect that the
possible values for the mass of the remnants are the critical masses associated to the extremal outermost horizons [or, in an equivalent way, to the minima of the function $M(y_\text{H})$].

\begin{figure}[b]
\includegraphics[width=7.5cm]{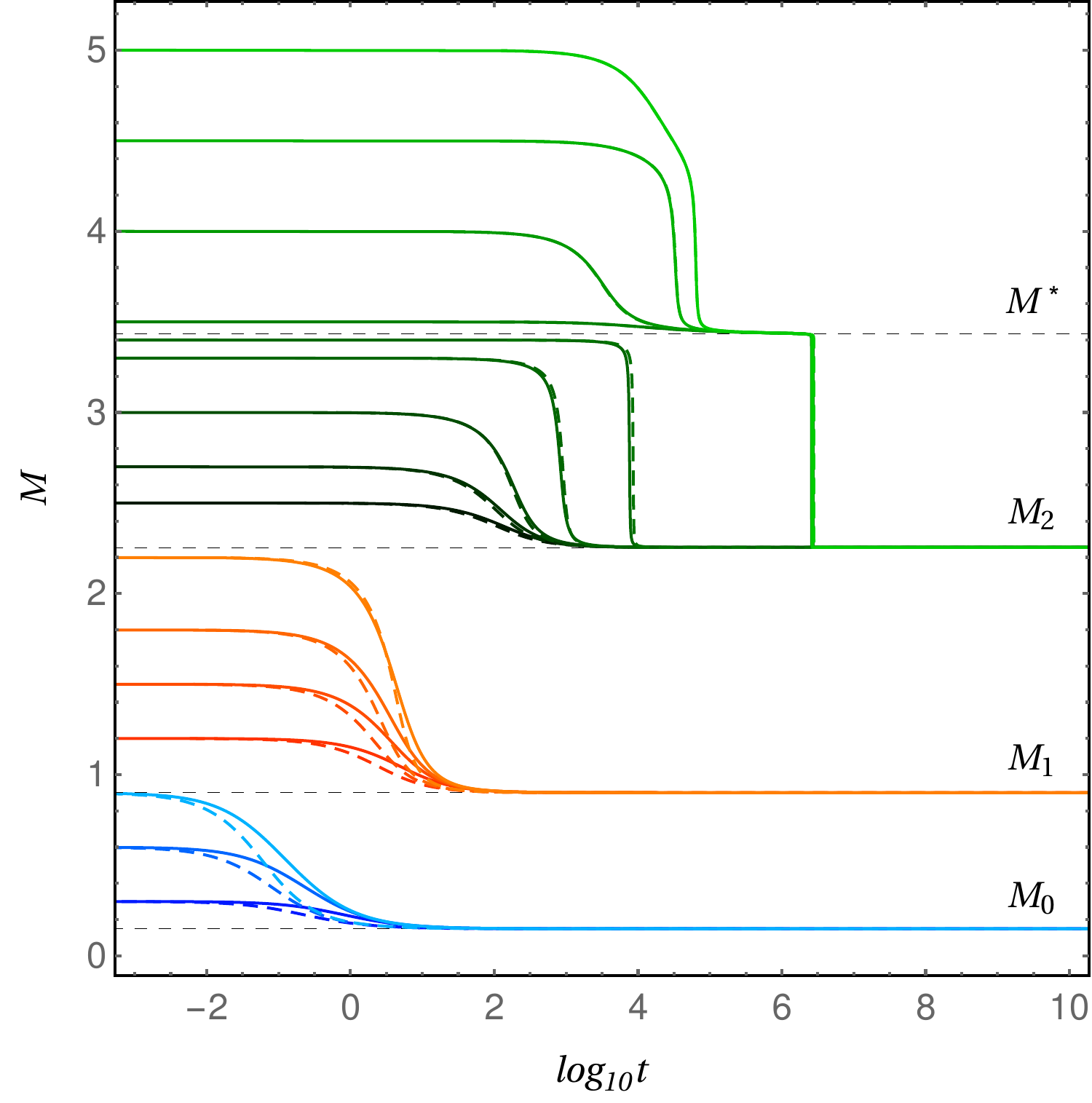}
\caption{\small  Time evolution of the total mass of the black hole (in units with $M_{\rm P}=1$) during evaporation,  obtained by the numerical integration of Eq.~\eqref{BHlifetime} in the case $a=1$, $q=3$ for different values of the initial mass $M_i = M(t=0)$. The solid lines represent the case $B(r)\equiv 0$, while the dashed lines correspond to the non-trivial $B(r)$ of Sec.~\ref{Sec3}.
The evaporation asymptotically reaches a zero-temperature state which mass depends on the initial mass of the black hole. In particular, the model with $q=3$ admits three black hole regimes, each of which has a critical mass $M_n$ (see Table~\ref{Tab4}). If the black hole is small (\textit{i.e.},  $M_{0} < M_i <M_{1}$) the evaporation will produce a remnant of mass $M_{0}= 0.149$ (blue curves); if it is an intermediate black hole with $M_{1}<M_i<M_{2}$ it will result in a remnant with $M_1=0.901$ (orange curves); finally, if the initial mass is larger than $M_{2}$ (green curves), the remnant will have mass $M_2=2.254$.
Since the function $Z_q(y)$ has a saddle point very close to $q=3$, there is a quasi-stable configuration with mass around $M^*  \approx 3.436$. This situation is related to the inflection point marked with a grey diamond in Fig.~\ref{Fig8c}, which also corresponds to a local minimum $T\approx 0$ of the Hawking temperature (see Fig.~\ref{Fig8f}).
}
\label{Fig9}
\end{figure}

An explicit example of this discussion is provided by Fig.~\ref{Fig9}, where we numerically solve the integral in~\eq{BHlifetime} for the Lee--Wick black holes with parameters $a=M_{\rm P}$, $q = 3$, for different values of initial mass. Notice that there are only three possibilities for the remnant's mass (indicated by the thin dashed lines). These values coincide with the ones listed in Table~\ref{Tab4} and correspond to the three possible black hole regimes for $q=3$. If the initial black hole has a mass between the critical masses $M_n$ and $M_{n+1}$, after evaporation it will result in an object with mass $M_n$.

Figure~\ref{Fig9} can also be used to exemplify another non-trivial feature of Lee--Wick black holes with $q>1.67$, that is the existence of quasi-stable configurations during the black hole evaporation. This can be viewed as the plateau that occurs for the light-green curves near $M^* \approx 3.436 M_{\rm P}$. The plateau exists because for $q=3.044$ the function $Z_q(y)$ has a saddle point at $y=20.148$ (see Table~\ref{Tab1}). Therefore, as we approach this saddle point, $A'(r_{\rm H}) \approx 0$ and the black hole is cold, \textit{i.e.}, $T \approx 0$ [see Eqs.~\eq{surgrav} and~\eq{T_au}], so that the evaporation becomes slow and it may take a long time to pass the regime with radius around $y=20.1$. This quasi-stable point where $M^{\prime\prime}(y_\text{H})=0$ and $M^\prime(y_\text{H}) \approx 0$ is marked with a grey diamond in Fig.~\ref{Fig8c}, and corresponds to a local minima of the Hawking temperature in Fig.~\ref{Fig8f}. In this case, the closest the initial mass $M_i$ is to $3.436 M_{\rm P}$, the longer is the lifetime of the intermediate state. 
In general, if $q$ is close to any of the $q^*$ such that $Z_{q^*}(y)$ has a saddle point for $y=y^*$, then besides the possible values of remnant masses, there might be a temporary state with $M^* \approx \frac{M_{\rm P}^2}{a \, Z_{q} (y^*)}$.


\section{Summary and conclusions}
\label{Sec6}

The only type of Lee--Wick black holes studied so far in the literature was the one associated with the action~\eq{action} with the parameter $\al_1=0$~\cite{Bambi:2016wmo}. As shown above, this choice considerably restricts the space of solutions as it fixes the real and imaginary parts of the Lee--Wick mass $\mu$ to have the same value.
In this vein, here we presented two families of Lee--Wick black holes (or regular horizonless objects). The first one, considered in Sec.~\ref{Sec2}, is a generalization of the solution obtained in~\cite{Bambi:2016wmo} to the whole space of parameters. It has a simple Schwarzschild-like form [Eq.~\eq{metricB=0}] with a regular core, but can exhibit a rich structure of horizons. The second class of solutions, introduced in Sec.~\ref{Sec3} [Eq.~\eq{metricBfinal}], is also a three-parameter family with a regular centre and exactly the same horizon configuration as the other solution. The advantage of this slightly more complicated metric is that it correctly matches the modified Newtonian-limit solution of the Lee--Wick theory~\eq{action} in the weak-field limit.

The largest portion of this work comprises the analysis of the structure of horizons, which is common to both classes of solutions and depends on three parameters, namely, two model parameters (the real and the imaginary parts of the Lee--Wick mass $\mu=a+ib$) and the mass $M$ of the source. Among these, the ratio $q=b/a$ is of utmost importance, as the most distinguished features of these theories are the oscillations of the effective delta source, mass function and gravitational potential --- which are all strongly dependent on $q$. For instance, while for $q < 1.67$ the metric can only describe either a horizonless compact object or a two-horizons black hole, for larger values of $q$ the solution can have a multiplicity of horizons; an estimate for the maximal number $N_\text{H}^{\text{max}}$ of horizons in terms of $q$ is provided by Eq.~\eq{HmaxAPP}. 

The actual number of horizons also depends on the value of the mass $M$ of the source. Simply put, if the mass is small enough the metric is horizonless; for intermediate values the metric can have two or more horizons, up to $N_\text{H}^{\text{max}}$; finally, for $M$ larger than a certain critical value the metric exhibits a fixed number of horizons that might be smaller than $N_\text{H}^{\text{max}}$. The horizons that remain in this large-$M$ limit are the ones related to the regions where the effective mass function $m(r)$ is negative, besides the trivial pair of horizons normally present in regular metrics. Since $m(r)$ only achieves negative values if $q > 2.67$, only beyond this threshold the metric can have more than two horizons for arbitrarily large values of $M$ --- in other words, if $q < 2.67$ the metric will have two horizons if $M \gg M_{\rm P}^2/a$.

The oscillations of the metric of general Lee--Wick black holes define not only a sequence of mass gaps for the number of horizons, but associated with it there is also a sequence of position gaps for the outermost horizon. This means that, depending on the parameters $a$ and $b$, black holes can only exist in certain specific regimes of radius. As a consequence of such structure, the final state of the evaporation of the black hole depends on which gap it was, and the possible values for the remnants' mass form a discrete set. In addition to these asymptotic states, if the parameter $q$ is close to a saddle point of the function $Z_q(y)$ there is the possibility of having a quasi-stable intermediate cold black hole configuration characterised by a longer evaporation lifetime.

The positions of the inner horizons are bounded to occur inside a sphere whose radius is inversely proportional to $b$, whereas there is no upper bound for the outer horizon. In fact, for large values of $M$ the event horizon tends to approach the Schwarzschild radius. In this sense, only if the parameters $a$ and $b$ of the model are sufficiently small (which is equivalent to having large parameters $\al_1$ and $\al_2$ in the action) the rich horizon structure can be extended to astrophysical scales. This situation is another manifestation of the weak seesaw-like mechanism in higher-derivative gravity, discussed in~\cite{Accioly:2016etf}. In the other extreme of the spectrum, such modifications of the Schwarzschild metric might be relevant for mini black holes. Even though the detection of these objects seems to be still beyond the current experimental facilities, their study is important from the theoretical side for the better understanding of the different approaches towards quantum gravity. Last but not least, the stability of regular black holes against perturbations is an important and active research topic (see, \textit{e.g.},~\cite{BonSauLiv} and references therein) and it would be interesting for future works to study if the solutions obtained here are stable.


\section*{Acknowledgments}

This work was supported by the Basic Research Program of the Science, Technology, and Innovation Commission of Shenzhen Municipality (grant no.\ JCYJ20180302174206969).


\appendix

\renewcommand{\thesubsection}{\thesection.\arabic{subsection}}

\section{Some results on the functions $\tilde{m}$ and $Z_q$}
\label{Appendix}

Here we prove some statements made in Sec.~\ref{Sec2} regarding the functions $\tilde{m}$ and $Z_q$, and show some properties of them.

\subsection{Maxima and minima of $\tilde{m}(y)$}

Let us elaborate more on the sequences of maxima and minima of the dimensionless mass function~\eq{mtil}, complementing the discussion at the end of Sec.~\ref{Sec2.massfunc}. From 
\beq
\label{mtilp}
\tilde{m}^\prime(y) = \frac{(1+q^2)^2}{2  q^3 }  e^{-\frac{y}{q}} y \sin y ,
\eeq
it follows that the maxima of $\tilde{m}(y)$ occur at $y=(2k-1)\pi$ and the minima, at $y=2k\pi$, with $k\in \lbrace 1,2,3\ldots\rbrace$. Substituting the values of the sine and cosine functions at these points back into~\eq{mtil}, it follows that the maxima and the minima of $\tilde{m}(y)$ are contained, respectively, in the curves $f_+(y)$ and $f_-(y)$, where
\beq
\label{fmassf}
f_\pm(y) = 1 \pm \frac{2 q + y (1 + q^2)}{2 q}e^{-\frac{y}{q}} .
\eeq
Since 
\beq
f_\pm'(y) = \pm \left[q \left(-1+q^2\right)-\left(1+q^2\right) y\right] \frac{e^{-\frac{y}{q}}}{2 q^2},
\eeq
the function $f_\pm(y)$ can only have at most one local stationary point, at
\beq
y = \frac{q \left(q^2-1\right)}{q^2 + 1}.
\eeq
Hence, $f_\pm(y)$ is monotonic if $0< q \leqslant 1$ (we always assume $y \geqslant 0$), while if $q>1$ the function $f_+(y)$ (respectively, $f_-(y)$) increases (decreases) up to a maximum (minimum), from where it decreases (increases) monotonically to zero. Nevertheless, since  $\tilde{m}(y)$ is equal to $f_\pm(y)$ only at the multiples of $\pi$, it can be verified that the range of values of $q$ for which the sequences of the extrema are monotonic is actually $0 < q < 6.07$, thus, larger than the above estimate. This explains why the amplitude of oscillations is decreasing in all the curves shown in Fig.~\ref{Fig1};
for larger values of $q$, however, the amplitude increases up to a maximum, before decreasing to zero.


\subsection{Position of the extrema of $Z_q(y)$}

Since the amplitude of the oscillations of $\tilde{m}(y)$ increases with $q$, we expect that (for $q$ sufficiently large) the position of the extrema of $Z_q(y)$ [see~\eq{Z}] should be close to those of $\tilde{m}(y)$, \textit{i.e.}, near multiples of $\pi$, as also suggested by Fig.~\ref{Fig2}. In fact, in the large-$q$ limit, the equation~\eq{Gzero} of the extrema approaches
\beq
\Big(\frac{y \cos y-\sin y}{y^2}+\sin{y} \Big) \,q^2 \approx 0 \, .
\eeq  
Thus, the positions $\tilde{y}_{k}$ of the extrema of $Z_q(y)$ are approximated by the solutions of
  \beq\label{SinCos}
  \frac{y \cos y-\sin y}{y^2}+\sin{y}=0 \, , \qquad y>0 ,
  \eeq
which can be obtained iteratively, namely, 
\beq\label{pertRootG}
\tilde{y}_{k} \approx k \pi -\frac{1}{k \pi} -\frac{5}{3 (k \pi)^3} +O(k^{-5}) \, , \qquad k = 1,2,\ldots .
\eeq
This verifies the statement that the extrema of $Z_q(y)$ tend to occur near multiples of $\pi$. 
Moreover, the vertical asymptotes~\eq{pertRootG} of the curve~\eq{Gzero} define gaps $\mathscr{J}_{k}\approx(\tilde{y}_{2k+1},\tilde{y}_{2k+2})\times \mathbb{R}^+$ (with $k\in\lbrace 0,1,2,\ldots\rbrace$) on the first quadrant of the $yq$-plane where the function $Z_q(y)$ has no extrema (see Fig.~\ref{Fig2}).


\subsection{Saddle points of $Z_q(y)$}

From the implicit function theorem applied to the function $\mathcal{G}(y,q)$ in~\eq{G}, in the context of Eq.~\eq{Gzero}, it follows that the saddle points $P_\ell = (y^*_\ell, q^*_\ell)$ (with $\ell = 1,2,\ldots$) defined in Sec.~\ref{Sec2.hor} satisfy
$\partial_y \mathcal{G} (y^*_\ell, q^*_\ell) = 0$,
which is equivalent to 
\beq
\left[ q y  \cos y + (q - y) \sin y\right]  \big\vert_{(y^*_\ell, q^*_\ell)} = 0 .
\eeq
For $q$ and $y$ sufficiently large, the term with the cosine dominates and the $y$-coordinates of the saddle points are, approximately, half-integer multiples of $\pi$. However, because of the existence of the regions $\mathscr{J}_{k}$ in the $yq$-plane where the curve $\mathcal{G}(y,q)=0$ is not defined,\footnote{See the discussion following Eq.~\eq{pertRootG}.} the saddle points also tend to have a periodicity of $2\pi$, \textit{i.e.},
\beq
\label{y*_app}
y^*_\ell \approx (4 \ell + 1) \frac{\pi}{2} , \qquad \ell=1,2,\ldots .
\eeq

Substituting the approximation~\eq{y*_app} back into~\eq{Gzero} and discarding the sub-leading terms in $q$ and $y$, we get
\beq
\label{appfur}
\left[-2  q^3 e^{\frac{y}{q}} + (1+q^2 )^2 y^2\right] \Big\vert_{(y^*_\ell, q^*_\ell)} \approx 0.
\eeq
If we also ignore the terms $O(q^0)$ and $O(q^2)$, the relation~\eq{appfur} can be solved for $q^*_\ell$, namely,
\beq
\label{Qstar}
q^*_\ell \approx \frac{y^*_\ell}{W \left(\frac{{y^*_\ell}^3}{2}\right)} \approx \frac{(4 \ell + 1) \pi }{2 \, W \left(\frac{\pi^3 (4 \ell + 1)^3}{16} \right)}   , \qquad \ell=1,2,\ldots , 
\eeq
where we used~\eq{y*_app}; here $W(x)$ is the Lambert function (product logarithm)~\cite{LambertW}. In Table~\ref{Tab3} we compare the approximation of the saddle points by formulas~\eq{y*_app} and~\eq{Qstar}  with the numerical values of Table~\ref{Tab1}. As expected, the errors $\De q^*_\ell$ and $\De y^*_\ell$ involved in the approximation decrease for larger values of $\ell$.

Moreover, regarding the right-hand side of~\eq{Qstar} as a function of $\ell$, it can be verified that this function is monotonically increasing if $\ell \geqslant 1$, what gives a more analytic support to the claim of Sec.~\ref{Sec2.hor} that the sequence $\lbrace P_\ell \rbrace_{\ell\in\mathbb{N}}$ of saddle points  is also ordered by increasing values of $q^*_\ell$.

\begin{table}[!ht]
    \centering
    \begin{tabular}{|r|r|r|}
    \hline
      $\ell$ & $\De q^*_\ell$ & $\De y^*_\ell$  \\
    \hline
    \hline
        1 & 0.25   & 0.43  \\ \hline
        2 & 0.15   & 0.33  \\ \hline
        3 & 0.10 	& 0.27   \\ \hline
        4 & 0.08 	& 0.23  \\ \hline
        5 & 0.07 	& 0.20   \\ \hline
        6 & 0.06 	& 0.18   \\ \hline
        7 & 0.05 	& 0.17  \\ \hline
        8 & 0.04 	& 0.15   \\ \hline
        9 & 0.04 	& 0.14   \\ \hline
       10 & 0.03 	& 0.13   \\ \hline
    \end{tabular}
    \caption{\small  
    Difference $\De q^*_\ell \equiv q^*_\ell\text{(approx.)} - q^*_\ell\text{(numerical)}$ and $\De y^*_\ell \equiv y^*_\ell\text{(approx.)} - y^*_\ell\text{(numerical)}$ between the approximate formulas~\eq{y*_app} and~\eq{Qstar} and the exact position of the first ten saddle points.}
    \label{Tab3}
\end{table}


\subsection{Number of extrema of $Z_q(y)$}

\begin{figure}[h]
\includegraphics[width=8cm]{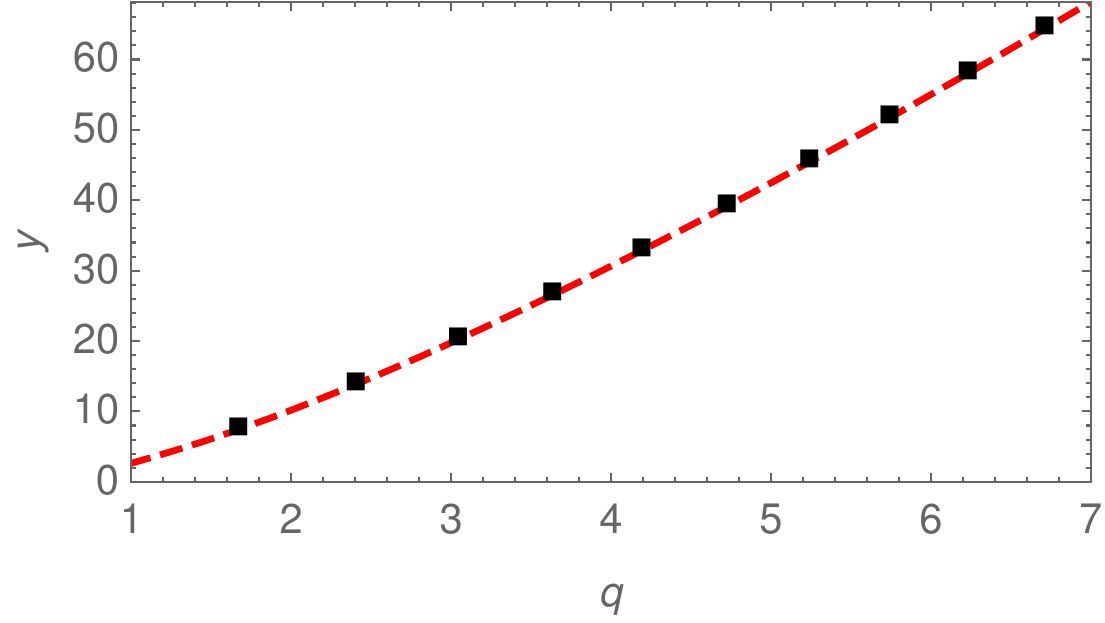}
\caption{\small   Comparison between \eqref{Ystar} (dashed line) and the numerical solutions $(q_\ell^{*},y_\ell^{*})$ of Table \ref{Tab1} (black squares).} \label{Fig4}
\end{figure}

It is possible to solve Eq.~\eq{appfur} for $y$, which results in an interpolation function $Y^*(q)$ between the approximate saddle points, namely, $Y^*(q^*_\ell) \approx y^*_\ell$ with
\beq
\label{Ystar}
Y^*(q) = -2q \, W_{-1} \left(-\frac{\sqrt{q}}{\sqrt{2}(q^2-1)}\right)   , \qquad q>1 \, ,
\eeq
where $W_{-1}(x)$ denotes the branch of the Lambert $W$ function such that $W(x) \leqslant -1$~\cite{LambertW}.
Notice that the function~\eqref{Ystar} is well defined and positive for $q>1$. Figure~\ref{Fig4} shows a comparison between \eqref{Ystar} and the exact values of the saddle points $(q_\ell^{*},y_\ell^{*})$ (see Table~\ref{Tab1}).

We can use $Y^*(q)$ combined with Eq.~\eq{y*_app} to estimate, given a certain $q$, the position of the closest saddle point and, by extension, the maximal number of horizons. Indeed, since the function $Z_q(y)$ (for a fixed $q$) has $n_{\text{E}}$ extrema occurring at $y$ around multiples of $\pi$, the next saddle point in the $yq$-plane is the one with
$\ell = (n_\text{E} + 1)/2$; thus, the comparison of~\eq{y*_app} with~\eq{Ystar} yields the estimate 
\beq
\label{NEAPP}
n_{\text{E}} (q) \approx 2 \bigg\lceil  \frac{Y^*(q)}{2\pi } - \frac14 \bigg\rceil -1  , \qquad q>1 \, ,
\eeq
where $\lceil x\rceil $ is the ceiling function, while for $q\leqslant 1$ we already know that $n_{\text{E}}=1$. From this formula and~\eq{Rel} it follows the estimate~\eq{HmaxAPP} of the maximal number of horizons.


\subsection{Suppression of oscillations and estimate of $M_\text{c}$}

Another application of the interpolating function $Y^*(q)$ is the estimation of the critical mass $M_\text{c}$ above which the number of horizons of the metric is fixed. This follows from the observation that the metric does not have any horizon for $y \gtrsim Y^*(q)$, which means that, beyond this value, the oscillations of $Z_q(y)$ are already completely damped, while those of $\tilde{m}(y)$ are significantly suppressed; therefore, we can take $\tilde{m}(y) \approx 1$ in this regime. Applying this approximation into~\eqref{Az} we obtain for the equation defining the event (outer) horizon,
\beq
\label{prosM}
\frac{2 G M a q }{y} \approx 1 .
\eeq
The importance of this expression is twofold. First, it shows that, in this regime, the position of the last  horizon is very close to the Schwarzschild radius $y_\text{S} =2GMaq$ (remember that $y=br=aqr$). Second, substituting the ``minimal'' value of $y$ which characterises the regime, provided by the interpolating function~\eq{Ystar}, and solving~\eq{prosM} for $M$ we obtain an estimate for $M_\text{c}$ as a function of $a$ and $q$, namely,
\beq
\label{EstMc}
M_\text{c}(a,q) \approx - \frac{M_{\rm P}^2}{a} \, W_{-1} \left(-\frac{\sqrt{q}}{\sqrt{2}(q^2-1)}\right)   , \qquad q>1 \, ,
\eeq
where $M_{\rm P}^2 = 1/G$ is the square of the Planck mass.

From the expansion of the Lambert function for small arguments~\cite{LambertW} 
it follows that the order of magnitude of $M_\text{c}$ is comparable to the one of $(\ln q) M_{\rm P}^2 /a$. Hence, if the real and imaginary parts of the Lee--Wick mass (\textit{i.e.}, $a$ and $b$) have similar orders of magnitude, for $M \gg M_{\rm P}^2 /a$ the number of horizons is fixed and the event horizon is very close to the Schwarzschild radius.


\section{Components of ${R^{\mu\nu}}_{\al\be}$ for the solution~\eq{metricB=0}}
\label{AppendixB}

The non-zero components of ${R^{\mu\nu}}_{\al\be}$ for a static spherically symmetric metric in the form~\eq{metricB}, with two arbitrary functions $A(r)$ and $B(r)$ are
\bea
\label{C1gen}
C_1 & \equiv & {R^{tr}}_{tr} = -\frac{A''(r)}{2}-\frac{1}{2} A(r) B''(r) -\frac{3}{4} A'(r) B'(r)-\frac{1}{4} A(r) B'^2(r)  ,
\\
\label{C2gen}
C_2 & \equiv & {R^{t\th}}_{t\th} = {R^{t\phi}}_{t\phi}  = -\frac{A'(r)+A(r) B'(r)}{2 r} ,
\\
\label{C3gen}
C_3 & \equiv & {R^{r\th}}_{r\th} = {R^{r\phi}}_{r\phi} =  -\frac{A'(r)}{2 r} ,
\\
\label{C4gen}
C_4 & \equiv & {R^{\th\phi}}_{\th\phi} = -\frac{A(r)-1}{r^2} .
\eea
Notice that $C_2 = C_3$ if $B(r) \equiv 0$. 
Therefore, for the metric~\eq{metricB=0} with $B(r) \equiv 0$ and $A(r)$ given by Eqs.~\eq{Am} and~\eq{meff}, it follows, 

\bea
C_1 & = & \frac{2 G M}{r^3} \left\lbrace 1 - \frac{e^{-a r}}{4 a b }  
\left[  b \left( 4 a + 2 c r - c^2 r^3  \right)  \cos(b r)
+ \left( 2 (a^2-b^2) + 2 a c r + c^2 r^2 + a c^2 r^3 \right)  \sin(b r) \right] \right\rbrace ,
\\
\label{C3B0}
C_3 &  = & -\frac{G M}{r^3} \left\lbrace 1 - \frac{e^{-a r} }{2 a b} \left[b \left(2 a + c r\right) \cos(b r) + \left(a^2-b^2 + a c r + c^2 r^2\right) \sin(b r)\right] \right\rbrace ,
\\
\label{C4B0}
C_4 & = &  \frac{2 G M}{r^3} \left\lbrace  1 - \frac{e^{-a r}}{2 a b } \left[b \left(2 a + c r \right) \cos(b r)+\left(a^2-b^2 + a c r\right) \sin(b r)\right] \right\rbrace ,
\eea
where we defined $c\equiv a^2 + b^2$. It is straightforward to verify that these functions are bounded everywhere; in particular, in the limit $r\to 0$ we get
\beq
\lim_{r\to 0} C_i  = \frac{(a^2+b^2)^2}{3 a}  G M , \qquad i=1,2,3,4.
\eeq

\section{Components of ${R^{\mu\nu}}_{\al\be}$ for the solution~\eq{metricBfinal}}
\label{AppendixC}

The non-zero components of ${R^{\mu\nu}}_{\al\be}$ for the solution~\eq{metricBfinal} can be calculated using Eqs.~\eq{C1gen}-\eq{C4gen}. Since the components $C_3$ and $C_4$ only depend on $A(r)$, and this function is the same for 
the metric~\eq{metricBfinal} and the one considered in Appendix~\ref{AppendixB}, it follows that $C_3$ and $C_4$ are given, respectively, by Eqs.~\eq{C3B0} and~\eq{C4B0}. The remaining $B(r)$-dependent components read
\beq
\begin{split}
C_1 & =  \frac{2G M}{r^3} - \frac{G M e^{-a r}}{2 a b r^3}  \left[2 b (2 a+c r) \cos(b r) + \left(2 a^2-2 b^2+2 a c r+c^2 r^2\right) \sin(b r)\right]
\\
& +\frac{c^2 (G M)^2 e^{-a r}}{2 a b r^2}  \left[ 2 b r \cos(b r)-(3+2 a r) \sin(b r)\right] 
+\frac{c^2 (G M)^2 e^{-2a r}}{8 a^2 b^2 r^2} \big[ -3 b^2+a \left(3 a+5 a^2 r-3 b^2 r+4 a c r^2\right)
\\
&
-\left(3 a^2-3 b^2+5 a c r+4 c^2 r^2\right) \cos(2 b r)+b (6 a+5 c r) \sin(2 b r) \big]
+\frac{c^4 (G M)^3 }{2 a^2 b^2 r} \left[ e^{-a r} \sin(b r)\right] ^2
\\
& -\frac{c^4 (G M)^3 e^{-a r}}{4 a^3 b^3 r} \left[ e^{-a r} \sin(b r)\right] ^2 \left[b (2 a+c r) \cos(b r)+\left(a^2-b^2+a c r\right) \sin(b r)\right] 
\end{split}
\eeq
and
\beq
C_2 = \frac{G M }{2 a^2 b^2 r^3} \left[ a b - G M c^2 r e^{-a r} \sin(b r) \right] \left\lbrace  -2 a b + e^{-a r} \left[  b (2 a+c r) \cos(b r) +  (a^2-b^2+a c r) \sin(b r) \right] \right\rbrace  ,
\eeq
where as before $c\equiv a^2 + b^2$. Like $C_3$ and $C_4$ (see Appendix~\ref{AppendixB}), 
these functions are finite everywhere. For instance, in the $r\to 0$ limit we have
\beq
\lim_{r\to 0} C_i  = -\frac{(a^2+b^2)^2}{6 a} G M , \qquad i=1,2.
\eeq


\end{document}